\documentclass[journal]{vgtc}                

\ifpdf
\pdfoutput=1\relax                   
  \usepackage{graphicx}                
\DeclareGraphicsExtensions{.pdf,.png,.jpg,.jpeg} 
\else
  \usepackage{graphicx}                
\DeclareGraphicsExtensions{.eps}     
\fi%

\graphicspath{{figures/}{pictures/}{images/}{./}} 

\usepackage{microtype}                 
\PassOptionsToPackage{warn}{textcomp}  
\usepackage{textcomp}                  
\usepackage{mathptmx}                  
\usepackage{times}                     
\usepackage{cite}                      
\usepackage{tabu}                      
\usepackage{booktabs}                  

\usepackage{amsmath,amsfonts}
\usepackage{algorithmic}
\usepackage{algorithm}
\usepackage{array}
\usepackage[caption=false,font=normalsize,labelfont=sf,textfont=sf]{subfig}
\usepackage{stfloats}
\usepackage{url}
\usepackage{verbatim}
\usepackage{graphicx}
\usepackage{bm}
\usepackage{makecell}
\usepackage{adjustbox}

\usepackage{multirow}
\newcolumntype{P}[1]{>{\centering\arraybackslash}m{#1}}
\usepackage[font=scriptsize]{caption}
\usepackage{hhline}

\usepackage{subfig}

\usepackage{xcolor}

\onlineid{0}

\vgtccategory{Research}
\vgtcpapertype{please specify}

\title{LFACon: Introducing Anglewise Attention to No-Reference Quality Assessment in Light Field Space}

\author{Qiang Qu, Xiaoming Chen, Yuk Ying Chung, \textit{Member, IEEE}, and Weidong Cai, \textit{Member, IEEE}}
\authorfooter{
\item
Qiang Qu, Yuk Ying Chung, Weidong Cai are with the University of Sydney, Australia.
\begin{sloppypar}E-mail: $\{vincent.qu \mid vera.chung \mid tom.cai\}@sydney.edu.au.$\end{sloppypar}
\item
Xiaoming Chen is with Beijing Technology and Business University, China.
\begin{sloppypar} E-mail: xiaoming.chen@btbu.edu.cn. \end{sloppypar}
\item
Xiaoming Chen is the corresponding author.
\item
Project URL: https://github.com/VincentQQu/LFACon
}

\shortauthortitle{Biv \MakeLowercase{\textit{et al.}}: Global Illumination for Fun and Profit}

\abstract{Light field imaging can capture both the intensity information and the direction information of light rays. It naturally enables a six-degrees-of-freedom viewing experience and deep user engagement in virtual reality. Compared to 2D image assessment, light field image quality assessment (LFIQA) needs to consider not only the image quality in the spatial domain but also the quality consistency in the angular domain. However, there is a lack of metrics to effectively reflect the angular consistency and thus the angular quality of a light field image (LFI). Furthermore, the existing LFIQA metrics suffer from high computational costs due to the excessive data volume of LFIs. In this paper, we propose a novel concept of ``anglewise attention'' by introducing a multihead self-attention mechanism to the angular domain of an LFI. This mechanism better reflects the LFI quality. In particular, we propose three new attention kernels, including anglewise self-attention, anglewise grid attention, and anglewise central attention. These attention kernels can realize angular self-attention, extract multiangled features globally or selectively, and reduce the computational cost of feature extraction. By effectively incorporating the proposed kernels, we further propose our light field attentional convolutional neural network (LFACon) as an LFIQA metric. Our experimental results show that the proposed LFACon metric significantly outperforms the state-of-the-art LFIQA metrics. For the majority of distortion types, LFACon attains the best performance with lower complexity and less computational time. %
} 


\CCScatlist{ 
\CCScat{K.6.1}{Management of Computing and Information Systems}%
{Project and People Management}{Life Cycle};
\CCScat{K.7.m}{The Computing Profession}{Miscellaneous}{Ethics}
}

\vgtcinsertpkg

\begin{document}


\firstsection{Introduction} \label{sec:intro}

\maketitle

Light field imaging is a cornerstone of immersive user experiences with six degrees of freedom in virtual reality, which enables the concurrent capture of multiple subviews (or subaperture images) \cite{meng20203d,koniaris2018compressed,date2019full}. Unlike traditional 2D imaging, light field imaging can record both the intensity and direction information of light rays in the real world \cite{wu2017light,itoh2016gaussian}. Thus, a light field image (LFI) has not only a spatial domain (inherited from 2D images) but also a unique angular domain. Each point in an LFI can be denoted as a 4D function $L(u,v,x,y)$, where $(u,v)$ represents the angular coordinate and $(x,y)$ represents the spatial coordinate.
Due to this distinct imaging principle, LFIs can naturally support a six-degrees-of-freedom viewing experience and user interaction. In LFI processing, such as compression and reconstruction, however, LFIs are prone to various types of distortions leading to degradation in the user perceived LFI quality. Therefore, in practical applications, it is vital to develop LFI quality assessment (LFIQA) metrics to quantitatively evaluate user perceived LFI quality and monitor a users' quality of experience.

\begin{figure}[htb]
\includegraphics[width=\columnwidth]{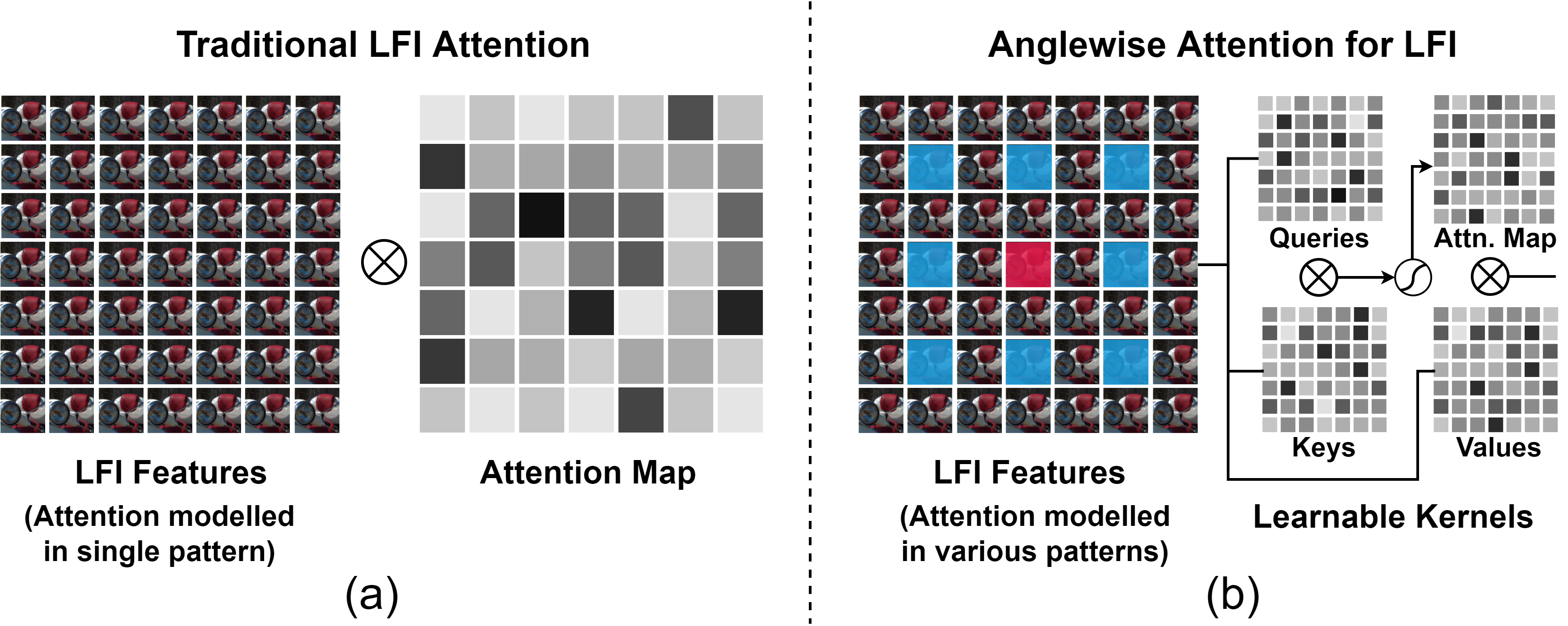}
\tiny
\caption{Different attention models for LFIs: (a) The existing LFI attention kernels directly multiply an LFI by a learned attention map. (b) The proposed kernels introduce learnable keys, queries, and values to better model the self-attention in various patterns (marked in different colors).}

\label{fig:attn_vs_asa}
\end{figure}

\begin{figure}[htb]
\includegraphics[width=\columnwidth]{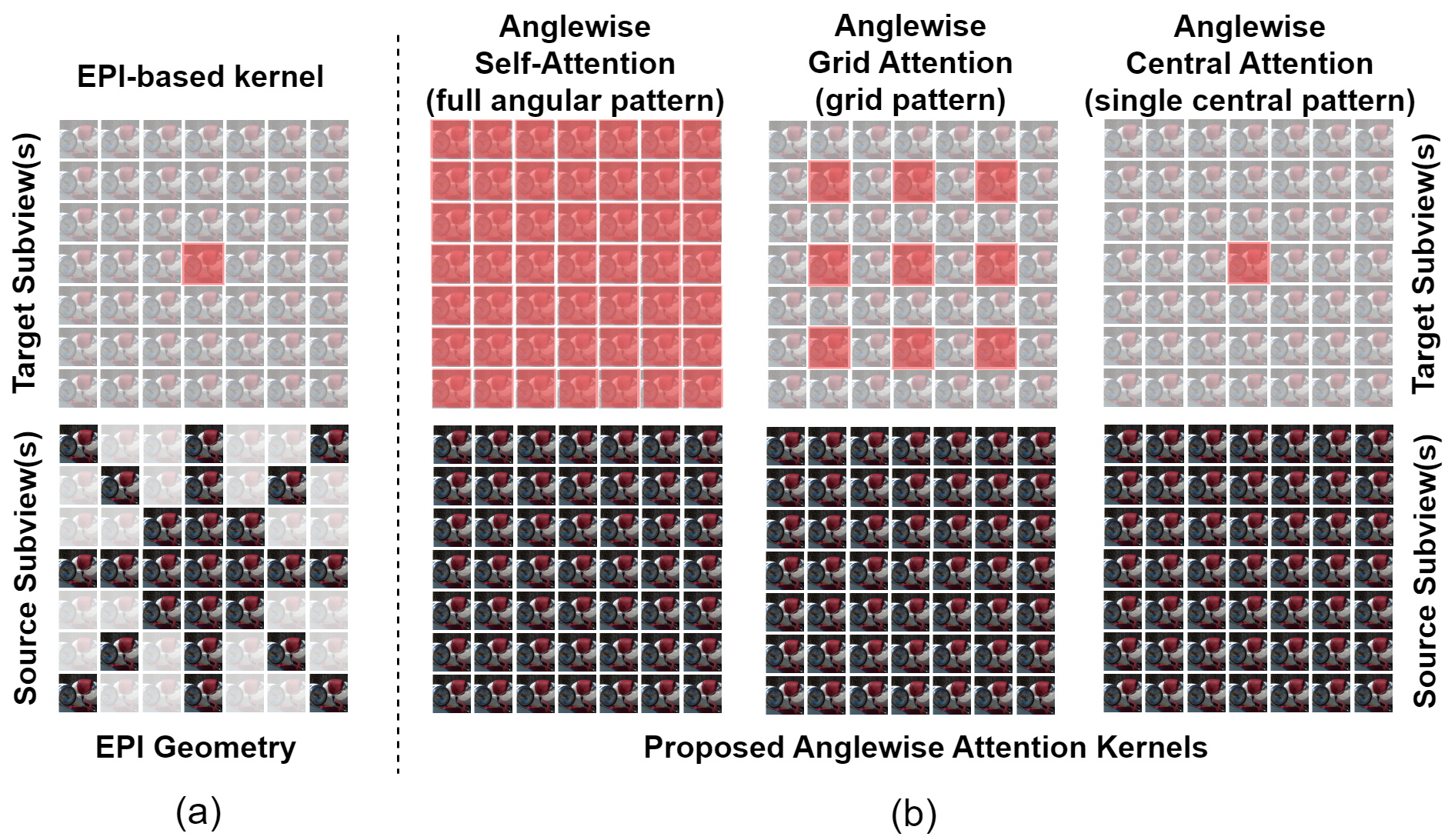}
\tiny
\caption{
Angular domain coverage in the attention computation: (a) The EPI-based kernels process the target subview by only utilizing partial source subviews (depending on the relative epipolar geometry). (b) The three proposed kernels utilize all the source subviews to compute the target subview's self-attention in three patterns with flexible complexity.}

\label{fig:epi_vs_asa}
\end{figure}

In the literature, LFIQA can be classified as a full-reference or no-reference task according to the availability of reference LFIs. Full-reference LFIQA assumes full access to the reference images in its quality assessment while no-reference LFIQA does not require any reference images \cite{shi2019no}. Therefore, no-reference LFIQA is of the utmost importance in real-world applications and is the primary focus of our research. However, no-reference LFIQA encounters several technical challenges.
First, it is challenging to leverage the correlations between different subviews to quantify the angular quality features of an LFI. In addition to the spatial quality, which can be assessed by traditional metrics, an LFIQA metric needs to reflect the angular quality, i.e., the quality consistency across subviews when a user views an LFI \cite{shi2019belif}.
Second, although attention mechanism is well studied in natural language processing \cite{luong2015effective, vaswani2017attention} and traditional computer vision \cite{zhang2019self, dosovitskiy2020image, liu2021swin}, it is difficult to model the unique ``angular attention'' in the light field. In fact, there is no effective self-attention mechanism that can model angular features from an LFI for quality assessment purposes. Such an attention mechanism, if developed, would enable selective concentration on key subviews for attention-aware feature extraction.
Third, due to the large data volume of an LFI, it is computationally expensive to perform feature extraction, which involves both the angular domain and the spatial domain. It is highly desirable to reduce this computational cost and to develop a lightweight metric for practical applications.

In response to the above challenges, we propose the novel concept of ``anglewise attention'' to improve no-reference LFIQA. Specifically, we propose three new anglewise attention kernels, including anglewise self-attention, anglewise grid attention, and anglewise central attention. Anglewise self-attention is used to create a self-attention mechanism in the angular domain and extract global multiangled features. On the other hand, anglewise grid/central attention is designed to selectively concentrate on the key subviews to extract the essential multiangled features with a lower computational cost.
The advantages of the proposed anglewise attention kernels for no-reference LFIQA are threefold.
First, the realized self-attention mechanism can better reflect the angular quality by considering the correlations between all the possible combinations of subview features.
Second, the proposed anglewise grid attention and anglewise central attention permit selective concentration on LFI features and alignment between the angular features across subviews, leading to improved feature extraction in the angular domain.
Third, all three anglewise attention kernels are open for flexible integration with other lightweight pre- and postattention kernels.
As a result, these attention kernels can combine the advantages of self-attention and the existing lightweight kernels for LFIs, e.g., the light field depthwise separable convolution (LF-DSC) proposed by Qu et al. \cite{qu2021light}, into a single generalized kernel. This can potentially lead to a more efficient no-reference LFIQA metric for practical use.

It is important to clarify the difference between the proposed anglewise attention kernels and the existing LFI attention methods. As shown in Fig. \ref{fig:attn_vs_asa}(a), the existing LFI attention mechanisms \cite{tsai2020attention,jin2020light,ma2022arfnet} directly multiply LFIs by a learned attention map. Our attention kernels, however, leverage the powerful multihead self-attention mechanism to better model self-attention in various patterns across subviews, as shown in Fig. \ref{fig:attn_vs_asa}(b). It is also worth noting that our attention kernels involve the full angular domain in the attention computation, which is different from the existing LFI processing techniques. For example, epipolar plane imaging (EPI) is one of the most widely used techniques in LFI processing \cite{shi2019belif, wanner2012globally}, but it only utilizes partial subviews according to the EPI geometry, as illustrated in Fig. \ref{fig:epi_vs_asa}(a). Our attention kernels, as shown in Fig. \ref{fig:epi_vs_asa}(b), can compute the multihead self-attention of different patterns with flexible complexity by utilizing all subviews (see Section \ref{sec:methodology} for details). As a result, the proposed kernels are expected to better reflect the angular quality consistency across all subviews.

Based on the proposed kernels, we design and implement a learning model, named the light field attentional convolutional neural network (LFACon), as an improved no-reference LFIQA metric. To evaluate the proposed metric, extensive experiments are conducted on three mainstream public datasets.
In our experiments, we first undertake an ablation study to determine the suitable pre- and postattention kernels for the proposed anglewise attention kernels. Then, extensive benchmarking experiments are conducted to compare the proposed metric with traditional image quality assessment metrics and state-of-the-art no-reference LFIQA metrics. The experimental results show that the proposed metric outperforms the state-of-the-art metrics and achieves the best performance for the majority of distortion types. In terms of computational cost, the proposed metric also exhibits comparatively shorter computation time as well as lower training costs.

The contributions of our work are summarized as follows:

\begin{itemize}
\item We propose the novel concept of anglewise attention and three new anglewise attention kernels for no-reference LFIQA. The proposed new kernels form a multihead self-attention mechanism to achieve efficient global or selective LFI feature extraction.

\item We conduct an ablation study to discover the optimal lightweight pre- and postattention kernels for the proposed attention kernels. The determined lightweight pre- and postattention kernels can be combined with any of the proposed attention kernels into a single generalized kernel, which can reduce the computational cost of the feature extraction.

\item We propose the LFACon metric, which incorporates an effective arrangement of the proposed anglewise attention kernels with the determined pre- and postkernels. This metric is verified by extensive experiments to show that it can significantly outperform the state-of-the-art no-reference LFIQA metrics.

\end{itemize}

\begin{figure*}[t]
\includegraphics[width=\textwidth]{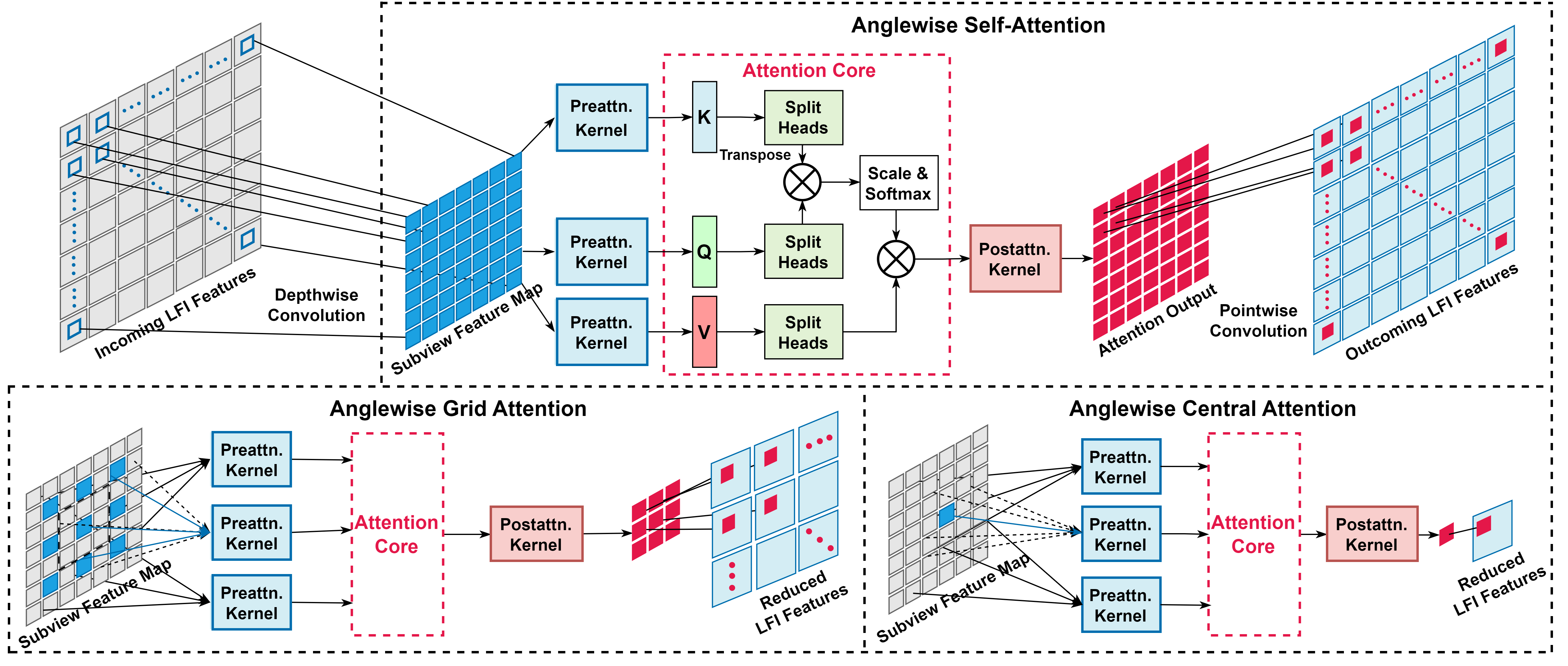}
\caption{Structure of the proposed anglewise attention kernels, including anglewise self-attention, anglewise grid attention, and anglewise central attention. The proposed anglewise attention kernels contain powerful multihead self-attention kernels and lightweight convolution operations (in pre- and postattention kernels), which are capsuled into a single generalized kernel for efficient feature extraction.}

\label{fig:kernel_overview}
\end{figure*}

\section{Related Work}
\label{sec:relatedworks}

\subsection{Light Field Imaging}
\label{subsec:light_field_imaging}

In contrast to conventional photography, which captures the two-dimensional projection of light rays, an LFI represents light rays from multiple directions. Due to the increased dimensionality of the data representation, light field photography enables the capture of more comprehensive visual data \cite{ihrke2016principles,yang2007toward,lv20204d}. The theoretical root of an LFI is interpreted by a plenoptic function, which is commonly represented as $l_{\lambda}(x,y,z,\theta,\phi, \lambda, t)$, where $l_{\lambda}[W/m2/sr/nm/s]$ denotes the spectral radiance per unit time and $(x,y,z)$ represents the spatial domain. Each point in a light field may be represented by a 4D coordinate $(u,v,x,y)$, where $(x,y)$ denotes the spatial domain and $(u,v)$ indicates the angular domain \cite{ihrke2016principles}. From the perspective of digital imaging, each pixel in an LFI can be located with two 2D coordinates, i.e., an angular coordinate identifying the subview in the LFI and a spatial coordinate indicating the exact pixel position within that subview, similar to a traditional 2D image.

\subsection{Attention Mechanism}
\label{subsec:attention_mechanism}

Self-attention can be used to learn the source material's alignments and has been shown to be very effective in natural language processing, especially after the introduction of the Transformer model \cite{vaswani2017attention}. In fact, self-attention has recently been shown to be advantageous in computer vision tasks. Zhang et al. \cite{zhang2019self} first introduced multihead self-attention into a 2D generative adversarial network. The Vision Transformer \cite{dosovitskiy2020image} and its variant with a shifted window, the Swin Transformer \cite{liu2021swin}, have been validated to outperform conventional convolution in a range of computer vision applications. However, these transformers require a much higher number of trainable parameters. The direct use of transformers on high-dimensional light field data is impractical owing to the astronomical training costs. This encourages us to introduce self-attention into the light field in a more appropriate way. Although there are several works that incorporate attention in the light field \cite{tsai2020attention,jin2020light,ma2022arfnet}, these approaches do not utilize powerful multihead self-attention. Fig. \ref{fig:attn_vs_asa} shows the difference between the existing LFI attention method and our proposed anglewise self-attention method. In addition, the existing approaches use attention and convolution independently. In contrast, our proposed anglewise attention is capable of combining the benefits of the attention mechanism and the lightweightness of separable convolution into a single generalized kernel (see Section \ref{subsec:aa} for details).

\subsection{Image Quality Assessment}
\label{subsec:iqa}
Objective image quality assessment methods can be categorized as full-reference metrics or no-reference metrics based on the availability of the original reference images \cite{shi2019no,huang2020light}. On the one hand, full-reference image quality assessment metrics assume full access to the reference images during the quality score prediction. On the other hand, no-reference metrics assess image quality without utilizing the original reference image, making them more challenging to implement but more applicable to real-world circumstances. Therefore, we focus on no-reference metrics in this research. On the one hand, numerous traditional metrics for 2D images have been proposed in the literature, including ERGAS \cite{wald2000quality}, UQI \cite{wang2002universal}, SSIM \cite{wang2004image}, multiscale SSIM \cite{wang2003multiscale}, RASE \cite{gonzalez2004fusion}, VSI \cite{zhang2014vsi}, VIF \cite{sheikh2006image}, PSNR-B \cite{yim2010quality}, and BRISQUE \cite{mittal2012no}. On the other hand, an increasing number of LFI-oriented metrics have been proposed. Huang et al. \cite{huang2020light} provided an overview of these metrics, including their classification, their evaluation criteria, and some publicly available LFI datasets. Recent research on no-reference LFIQA includes BELIF \cite{shi2019belif}, NR-LFQA \cite{shi2019no}, Tensor-NLFQ \cite{zhou2020tensor}, and ALAS-DADS \cite{qu2021light}. However, these metrics cannot sufficiently reflect LFI quality by taking both spatial quality and angular consistency into account in a comprehensive way. Nonetheless, in ALAS-DADS, Qu et al. \cite{qu2021light} extended depthwise separable convolution (DSC) to the spatial and angular domains of an LFI and introduced the light field version of DSC, i.e., LF-DSC, which is capable of extracting both spatial and angular features with lower computational cost. In this work, we consider LF-DSC as a promising candidate for the pre- and postattention kernels in our proposed anglewise attention kernels (discussed and reported in Section \ref{sec:experiments}).

\begin{figure*}[t]
\includegraphics[width=\textwidth]{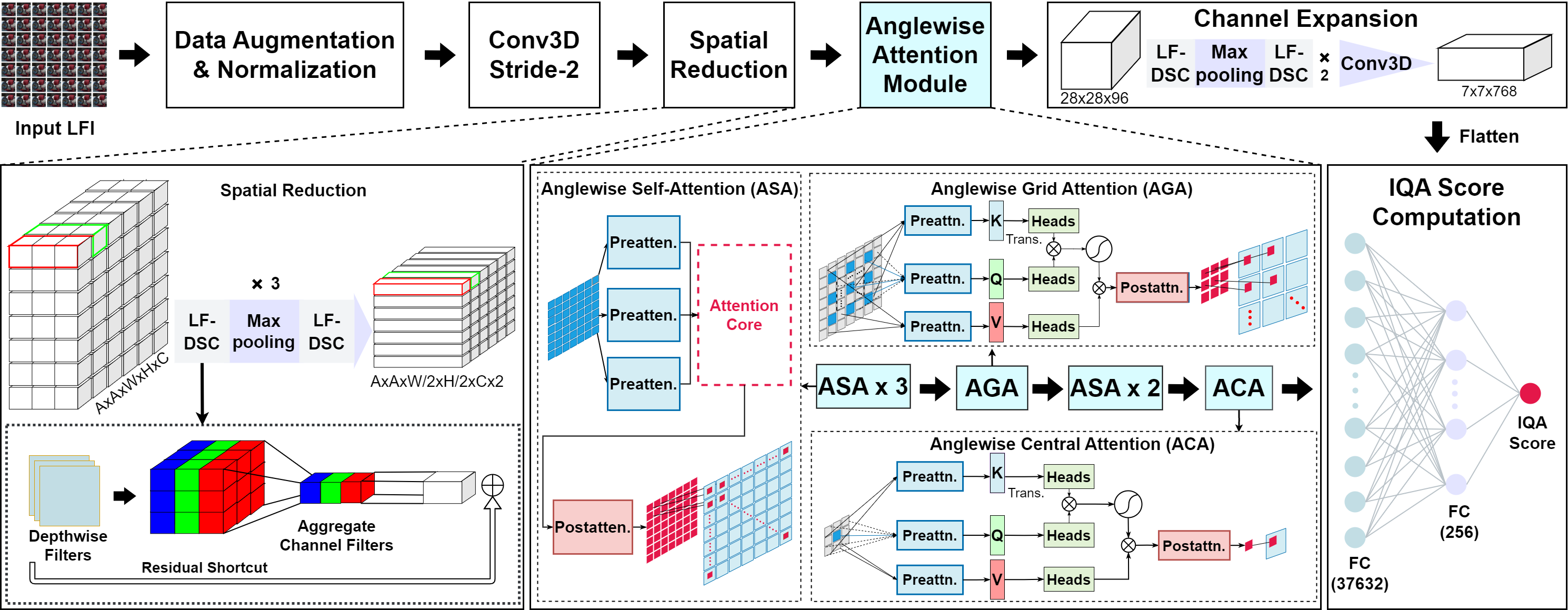}
\caption{Structure of the proposed LFACon metric, which incorporates the proposed anglewise attention kernels.}
\label{fig:model_overview}
\end{figure*}

\section{Methodology}
\label{sec:methodology}

Self-attention has recently been demonstrated to be beneficial in 2D image processing \cite{dosovitskiy2020image, liu2021swin}. In this work, we expand the scope of self-attention to the angular domain of LFIs and propose, for the first time, the novel concept of anglewise attention. Moreover, we propose three new anglewise attention kernels, including anglewise self-attention, anglewise grid attention, and anglewise central attention. Anglewise self-attention is designed to extract the multiangled features globally, while anglewise grid attention and anglewise central attention are devised to achieve selective feature concentration. On the basis of the proposed new kernels, we propose a learning model, LFACon, as an improved no-reference LFIQA metric.

\subsection{Anglewise Attention}
\label{subsec:aa}
As mentioned in Section \ref{subsec:aa}, unlike previous attention-based kernels, in essence, the proposed anglewise attention kernels leverage multihead self-attention to extract the multiangled features from the LFI. These kernels are designed to selectively emphasize the essential proportions of the LFI features, acquire the alignments between the diverse angular subview features and reduce the dimension in feature extraction. To be more specific, the three proposed kernels are introduced in detail as follows.

\subsubsection{Anglewise Self-Attention}
\label{subsec:asa}
Fig. \ref{fig:kernel_overview} illustrates the design of the proposed anglewise self-attention model. Assuming that we have incoming 5D LFI features $LF \in \mathbb{R}^{u\times v \times x \times y \times c}$ (two angular dimensions $u\times v$, two spatial dimensions $ x \times y$, and one channel dimension $c$), 
we first generate a subview feature map from these features with a depthwise convolution operation. This feature map is then transformed into three spaces $K, Q, V \in \mathbb{R}^{u\times v \times x \times y \times c}$, i.e., keys, queries, and values. The transformation is performed via three independent learnable preattention kernels $PreAttn_{i}(LF), \; i \in \{k,q,v\}$. Second, suitable pre- and postattention kernels for anglewise self-attention need to be determined. According to the related research \cite{vaswani2017attention, zhang2019self}, intuitively, the candidate pre- and postattention kernels can be either linear neural projections or $1 \times 1$ convolutional layers. However, since the recently proposed lightweight LF-DSC kernel \cite{qu2021light} is apparently an efficient and low-complexity feature extractor for LFIs, we propose LF-DSC as an additional promising candidate for the pre- and postattention kernels. To evaluate these candidates, we conduct an ablation study that confirms the superior performance of LF-DSC (the results are presented in Section \ref{sec:experiments}). Third, we flatten the spatial channel dimensions of $K, Q, V$ (s.t. $K, Q, V \in \mathbb{R}^{A \times d}$) before computing the attention value. Fourth, we split queries, keys, and values into multiple heads
and apply the attention function, respectively. The split multiple heads allow anglewise self-attention to cooperatively focus on the perceptual features from the distinct representation subspaces of different subviews, hence extending the kernel's capacity to take into account a wider range of subviews (instead of overfocusing on the neighbor subviews only). Finally, we compute the attention output by applying the scaled dot-product of the keys and queries, followed by a postattention kernel, i.e., $PostAttn(X)$.

Suppose we apply attention with $h$ heads, e.g., $h=8$. For each head $H_j\,, \; j \in [0, h-1]$:\\
\addtolength{\jot}{10pt}
\begin{equation}
\label{eq:asa_head}
    H_j = softmax(\frac{QK^T}{\sqrt{d}})V,
\end{equation}
where\\
\begin{equation}
\label{eq:asa_kqv}
\begin{split}
    K, Q, V = Z_i = PreAttn_{i}(LF), \quad i \in \{k,q,v\}, \\
    Z_i \in \mathbb{R}^{A \times d}, \;\; A = u\times v,\; d=\frac{x \times y \times c}{h}.
\end{split}
\end{equation}
The attention output is calculated as: \\
\begin{equation}
\label{eq:asa_out}
    Output_{Attn} = PostAttn(Concat(H_1, ... H_h)).
\end{equation}
$Output_{Attn}$ is then subjected to a pointwise convolution for a linear projection to yield the outcoming LFI features.

\subsubsection{Anglewise Grid Attention}
\label{subsec:aga}
In light field processing, selective concentration on the key features is often a crucial issue. This is because, in practice, light field processing tasks require output in significantly lower dimensions, e.g., 2D output for depth estimation \cite{chen2018accurate}, 1D output for classification \cite{lu2019improved}, and 1D output for LFIQA \cite{shi2019belif}.
To this end, anglewise grid attention is proposed to extract the multiangled features from a reduced dimension. Instead of applying self-attention globally to the entire angular domain (as anglewise self-attention does), anglewise grid attention only applies selective attention to an essential subset of angular features in a grid pattern.
We design this grid-based attention model over global attention because, in practice, the majority of the angular information can be derived from the disparity of neighboring angular subview features. As illustrated in Fig. \ref{fig:kernel_overview}, only the most significant angular features in the grid pattern are chosen because they can reasonably characterize the entire angular domain.

Accordingly, the computation of the attention heads of anglewise grid attention is different from equations \ref{eq:asa_head} \& \ref{eq:asa_kqv} in anglewise self-attention. Suppose we apply anglewise grid attention with a $b \times b$ neighborhood window set $W$ and a stride $s$. For each local window $w \in W$, $g$ denotes the central subview feature. Then, each local aggregated attention head $H_m$ out of $h$ heads (i.e., $m \in [0, h-1]$) is:

\begin{equation}
\label{eq:aga_head}
    H_m = Concat(H^{w}) \quad \forall w \in W, \\
\end{equation}
\begin{equation}
    H^w = softmax(\frac{Q^w {K^w}^T}{\sqrt{d}})V^w,
\end{equation}
where\\
\begin{equation}
\label{eq:aga_kqv}
    K^w, V^w = PreAttn_{i}(w), \quad i \in \{k,v\}, \\
\end{equation}
\begin{equation}
    Q^w = PreAttn_{q}(g).
\end{equation}
The remaining processing of anglewise grid attention is almost identical to that of anglewise self-attention, except that anglewise grid attention's outputs have significantly reduced dimensions.

\subsubsection{Anglewise Central Attention}
\label{subsec:aca}
As illustrated in Fig. \ref{fig:kernel_overview}, the design of anglewise central attention is inherited from anglewise grid attention but with further feature concentration. Instead of selecting the angular features from multiple subviews in a grid pattern, we choose the feature in the central subview and apply global self-attention to approximate the whole angular domain. The equations for anglewise central attention are the same as those for anglewise grid attention, except that the queries are at the center of the queries $Q$ in anglewise self-attention.

\subsection{LFACon Metric}
\label{subsec:learning_framework}
Based on the presented new kernels, we propose a new learning model, LFACon, as an improved no-reference LFIQA metric. Fig. \ref{fig:model_overview} shows the structure of LFACon, and described as follows.

\begin{table}[htbp]
\caption{RMSE $\downarrow$ comparison of multiple combinations of the candidate pre- and postattention kernels (best results in bold).}
\label{tab:kernels}
\scriptsize
\renewcommand{\arraystretch}{1.2}
\setlength{\tabcolsep}{2pt}
\begin{tabular}{P{0.14\textwidth}|P{0.07\textwidth}|P{0.07\textwidth}|P{0.07\textwidth}|P{0.09\textwidth}}
\hline
\hline
Kernel & Win5-LID & SMART &  MPI-LFA & Param. Level\\
\hline
Linear + Linear  & $0.4558$&  $0.8853$& $0.8714$ &  High\\
Conv3D + Conv3D  & $0.3972$&  $0.7960$ & $0.8209$ & Low\\
\textbf{Conv3D + LF-DSC}  & $\bm{0.3403}$&  $0.7206$ &  $\bm{0.7539}$ & Low\\
LF-DSC + LF-DSC & $0.3630$&  $\bm{0.7027}$ & $0.7631$ &  Medium \\
\hline
\hline
\end{tabular}
\end{table}

First, we perform normalization within the training set. For an LFI $k$,
\begin{equation} \hat{p}_k(D)=\frac{p_k(D)-\mu(D)}{\sigma(D)+1}, \quad \forall D=(u,v,x,y,c), \label{eq:norm}\end{equation} where $(u, v)$ is used to identify a subview and $(x, y, c)$ denotes the spatial coordinate and the color channel of a pixel in the subview.
Second, we reduce the spatial dimension of the model input by performing stride-2 3D convolution operations followed by a combination of 3D convolution operations and three repetitions of ``LF-DSC + max pooling + LF-DSC'' operations \cite{qu2021light}. The combined operations rapidly reduce the spatial dimensions for spatial feature extraction. After the combined operations, the spatial dimension is lowered from $434 \times 434$ to $28 \times 28$.
Third, anglewise attention is computed with a verified arrangement of our proposed attention kernels (the results are shown in Section \ref{sec:experiments}). In this process, the spatially reduced LFI is initially processed by three anglewise self-attention kernels to extract the angular features. The extracted features are then processed by one anglewise grid attention kernel to concentrate the angular features while reducing the angular dimensions from $7 \times 7$ to $3 \times 3$, followed by another two anglewise self-attention kernels.
Then, the angular dimension is further reduced to $1 \times 1$ with an anglewise central attention kernel.
Finally, we apply channel expansion to increase the number of channels from $96$ to $384$ for the following image quality score prediction, followed by a fully connected neural network to output the final quality scores.
Notably, at the end of each LF-DSC step, we choose layer normalization over batch normalization because layer normalization does not presume large batch sizes \cite{ba2016layer}. This resolves the issue of small batch sizes in LFI normalization due to its high dimensionality. In addition, we choose a smooth activation function, called Swish, due to its demonstrated capability in avoiding preactivation functions \cite{ramachandran2017searching}:

\begin{equation}
  \label{eq:swish}
  Swish(x) = \frac{x}{1 + e^{-x}}.
\end{equation}

Finally, the mean squared error (MSE) is leveraged as the loss function for training.

\begin{table*}[htb]
\caption{Quantitative comparisons (RMSE/SRCC/PLCC/OR) of the benchmarked image quality assessment metrics on the Win5-LID, SMART, and MPI-LFA datasets. The best results are in bold and the last row measures the improvement compared to the second-best result.}
\label{tab:benchmarking}
\scriptsize
\renewcommand{\arraystretch}{1.2}
\setlength{\tabcolsep}{2pt}
\begin{tabular}{P{0.06\textwidth}|P{0.125\textwidth}|P{0.058\textwidth}P{0.058\textwidth}P{0.058\textwidth}P{0.058\textwidth}|P{0.058\textwidth}P{0.058\textwidth}P{0.058\textwidth}P{0.058\textwidth}|P{0.058\textwidth}P{0.058\textwidth}P{0.058\textwidth}P{0.058\textwidth}}
\hline
\hline
&&\multicolumn{4}{c}{Win5-LID}&\multicolumn{4}{c}{SMART}&\multicolumn{4}{c}{MPI-LFA}\\
\hline
Type & Metrics & RMSE  $\downarrow$ & SRCC $\uparrow$ & PLCC $\uparrow$& OR  $\downarrow$ & RMSE  $\downarrow$& SRCC $\uparrow$ & PLCC $\uparrow$ &OR  $\downarrow$ & RMSE  $\downarrow$ & SRCC $\uparrow$ & PLCC $\uparrow$ &OR  $\downarrow$ \\
\hline
\multirow{12}{*}{Full Ref.}&PSNR & $0.7770$ & $0.6579$ & $0.6622$ & $0.0028$ & $1.8047$ & $0.6559$ & $0.5825$ & $0.0049$ & $1.8207$ & $0.4191$ & $0.3945$ & $0.0633$ \\
&PSNR-B & $0.7883$ & $0.6467$ & $0.6496$ & $0.0028$ & $1.8084$ & $0.6614$ & $0.5802$ & $0.0000$ & $1.8301$ & $0.4012$ & $0.3833$ & $0.0000$ \\
&SSIM & $0.8202$ & $0.5970$ & $0.6117$ & $0.0170$ & $1.8640$ & $0.5183$ & $0.5433$ & $0.0171$ & $1.9160$ & $0.4025$ & $0.2549$ & $0.1024$ \\
&MS-SSIM & $0.8036$ & $0.6350$ & $0.6320$ & $0.0568$ & $1.7955$ & $0.5317$ & $0.5882$ & $0.0318$ & $1.9275$ & $0.4451$ & $0.2316$ & $0.1304$ \\
&MSE & $0.8644$ & $0.6579$ & $0.5522$ & $0.0483$ & $1.8478$ & $0.6559$ & $0.5544$ & $0.0465$ & $1.9805$ & $0.4191$ & $0.0300$ & $0.1378$ \\
&RMSE-SW & $0.8436$ & $0.6128$ & $0.5814$ & $0.0312$ & $1.8522$ & $0.5902$ & $0.5514$ & $0.0196$ & $1.9716$ & $0.3860$ & $0.0996$ & $0.0819$ \\
&RASE & $0.9272$ & $0.4608$ & $0.4475$ & $0.0057$ & $2.1805$ & $0.1370$ & $0.1884$ & $0.0954$ & $1.9770$ & $0.3415$ & $0.0668$ & $0.1155$ \\
&SAM & $0.8528$ & $0.5904$ & $0.5687$ & $0.0142$ & $2.0123$ & $0.4422$ & $0.4226$ & $0.0147$ & $1.9599$ & $0.4093$ & $0.1468$ & $0.0875$ \\
&SCC & $0.8504$ & $0.5532$ & $0.5722$ & $0.0000$ & $2.1430$ & $0.2383$ & $0.2615$ & $0.0098$ & $1.8280$ & $0.3716$ & $0.3858$ & $0.0708$ \\
&ERGAS & $0.9262$ & $0.4633$ & $0.4495$ & $0.0057$ & $2.0759$ & $0.2218$ & $0.3546$ & $0.0269$ & $1.9768$ & $0.3419$ & $0.0684$ & $0.1117$ \\
&UQI & $0.9287$ & $0.4865$ & $0.4446$ & $0.0597$ & $2.0871$ & $0.1944$ & $0.3412$ & $0.0465$ & $1.9785$ & $0.3512$ & $0.0544$ & $0.1322$ \\
&VIF & $0.7970$ & $0.6241$ & $0.6396$ & $0.0000$ & $2.0456$ & $0.4049$ & $0.3888$ & $0.0000$ & $1.7479$ & $0.4156$ & $0.4709$ & $0.0000$ \\
\hline
\multirow{5}{*}{No Ref.}&BRISQUE & $0.5816$ & $0.8279$ & $0.8242$ & $0.0000$ & $1.2250$ & $0.7680$ & $0.8229$ & $0.0000$ & $1.4433$ & $0.5503$ & $0.6691$ & $0.0000$ \\
&NR-LFQA & $0.5638$ & $0.9258$ & $0.8746$ & $0.0000$ & $0.8689$ & $0.8884$ & $0.9213$ & $0.0000$ & $1.1302$ & $0.6954$ & $0.8215$ & $0.0000$ \\
&Tensor-NLFQ & $0.4482$ & $0.9008$ & $0.9075$ & $0.0000$ & $1.1997$ & $0.8383$ & $0.8489$ & $0.0000$ & $1.2829$ & $0.6425$ & $0.7618$ & $0.0000$ \\
&ALAS-DADS & $0.4371$ & $0.9047$ & $0.9068$ & $0.0000$ & $0.9749$ & $0.8480$ & $0.8984$ & $0.0000$ & $1.2987$ & $0.7000$ & $0.7552$ & $0.0000$ \\
&\textbf{Proposed LFACon} & $\bm{0.3403}$ & $\bm{0.9494}$ & $\bm{0.9518}$ & $\bm{0.0000}$ & $\bm{0.7206}$ & $\bm{0.8907}$ & $\bm{0.9508}$ & $\bm{0.0000}$ & $\bm{0.7539}$ & $\bm{0.8262}$ & $\bm{0.9335}$ & $\bm{0.0000}$ \\
\hline
&\textbf{Boost v.s. 2nd best} & $\bm{+28\%}$ & $\bm{+0.024}$ & $\bm{+0.044}$ & $\bm{-}$ & $\bm{+21\%}$ & $\bm{+0.002}$ & $\bm{+0.030}$ & $\bm{-}$ & $\bm{+50\%}$ & $\bm{+0.126}$ & $\bm{+0.112}$ & $\bm{-}$ \\
\hline
\hline
\end{tabular}
\end{table*}

\begin{figure*}[htbp]
\centering
\includegraphics[width=\linewidth]{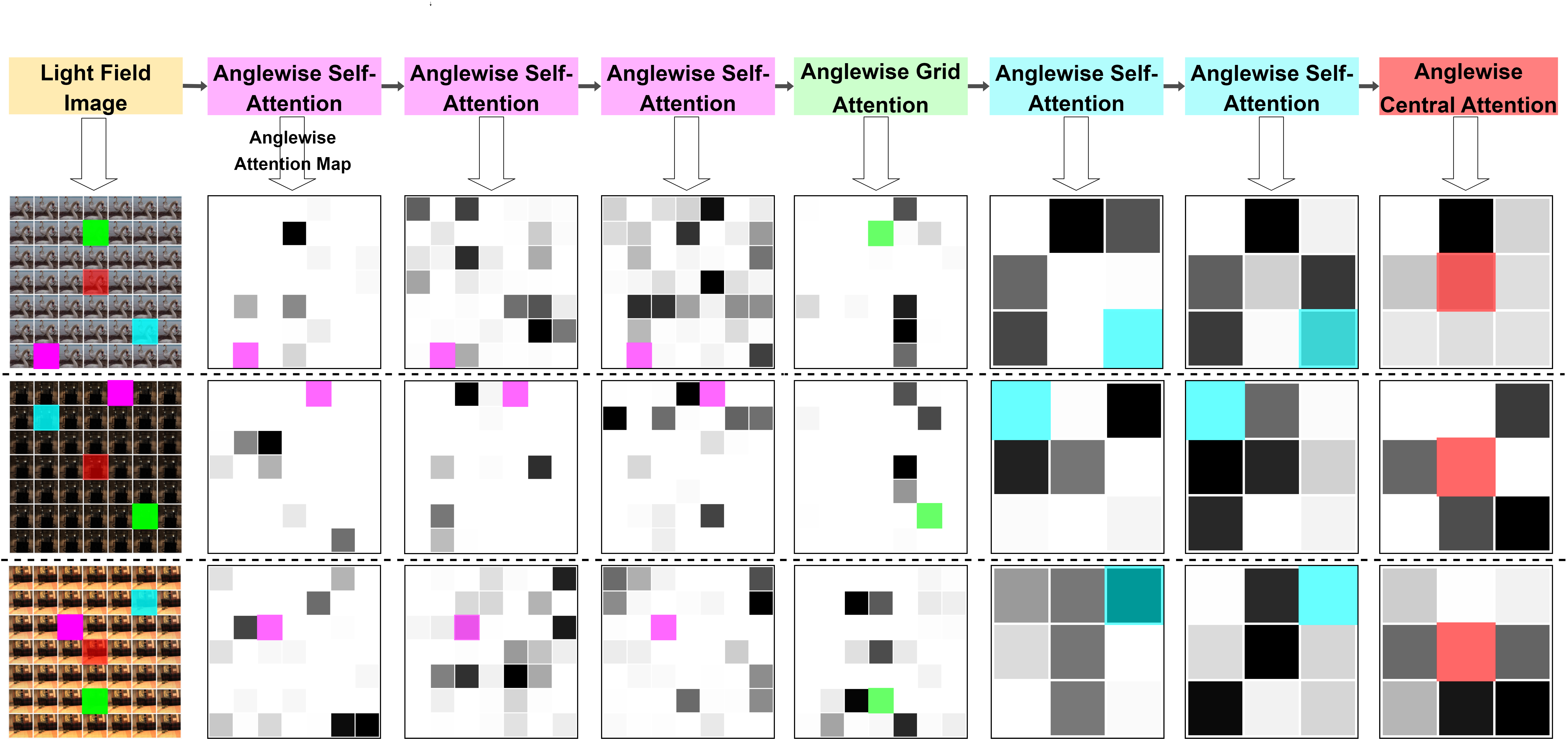}
\caption{Visualization of the proposed anglewise attention: The workflows and attention maps for the incorporated attention kernels in LFACon for three sample LFIs are shown in the three rows. In each row, the first column shows the sample LFI with four representative query locations marked in red, pink, blue, and green. The remaining seven columns show the anglewise attention maps (computed at different stages) for those query locations. The colored locations correspond to the query locations. The other locations are marked with different grayscale levels to visualize the different attention weights.}
\label{fig:attn_maps}
\end{figure*}

\section{Experiments}
\label{sec:experiments}

\subsection{Experimental Setup}
To verify the efficacy of the proposed attention kernels and the proposed LFACon metric, comprehensive experiments are conducted on three publicly accessible mainstream LFI datasets, including Win5-LID \cite{shi2018perceptual}, SMART \cite{paudyal2017towards}, and MPI-LFA \cite{kiran2017towards}. According to Yeung et al. \cite{yeung2018fast}, we resize and reshape all the LFIs to the same size of $7\times 7\times 434\times 434\times 3$ to retain the most relevant image portions.

\subsubsection{Datasets}
The Win5-LID dataset contains 6 real scenes and 4 synthetic scenes as reference LFIs. Then, six distortion types and five distortion levels were applied to the 10 reference LFIs, yielding 220 distorted LFIs. The six distortion types were HEVC, JPEG 2000, linear interpolation (LN), nearest-neighbor (NN) interpolation, EPICNN, and USCD. Participants assessed the quality of these 220 LFIs on a 5-point discrete scale using a double-stimulus continuous quality measure. For each LFI, the mean opinion scores were collected. The SMART dataset was built on 16 original LFIs, and 256 LFIs were warped by four distortion types, namely, HEVC Intra, JPEG, JPEG 2000, and SSDC. The subjective assessments were collected using the Bradley-Terry scoring system. The MPI-LFA dataset comprises fourteen immaculate LFIs. The 336 distorted LFIs were generated by applying six types of distortions, including HEVC, DQ, OPT, LINEAR, NN, and GAUSS. To evaluate the quality of the LFI, the pairwise comparison approach with a two-alternative forced choice was employed. The just objectionable difference value was recorded as the indicator of quality.

The ground truth (human visual perception) of the three datasets warrants further description. The quality ratings of the images in the Win5-LID dataset consist of the mean opinion scores ranging from 1 to 5, with higher values signifying better quality. In contrast, the SMART dataset uses Bradley-Terry scoring, which normally goes from -13 to 0, with larger numbers signifying better quality. The quality indicator used for the MPI-LFA dataset is the just objectionable difference score, which normally ranges from -9 to 0, with larger values indicating better quality.

\begin{figure}[htb]

\scriptsize
\renewcommand{\arraystretch}{1.2}
\setlength{\tabcolsep}{2pt}
\begin{tabular}{P{0.20\columnwidth}|P{0.18\columnwidth}|P{0.18\columnwidth}|P{0.18\columnwidth}|P{0.18\columnwidth}} 
\hline
\hline

\multicolumn{5}{c}{\textbf{Greek (JPEG2000 distortion)} from Win5-LID}\\
\hline
\hline

\multicolumn{5}{c} {\adjustbox{valign=c,margin=0 1pt 0 1pt}{\includegraphics[width=0.98\columnwidth, height=0.28\columnwidth]{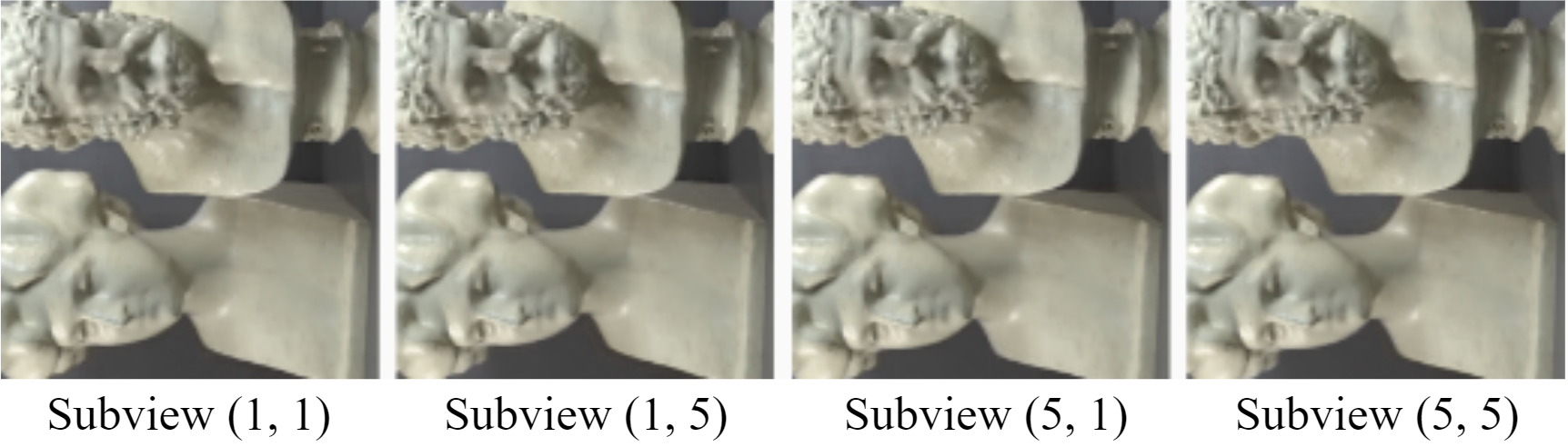}} }\\
\hline

\multicolumn{5}{c}{Ground Truth Quality Score: \underline{1.8696}  (in [1, 5])}\\
\hline
Metric & NR-LFQA & Tensor-NLFQ & ALAS-DADS & \textbf{LFACon}\\   
\hline
Quality Score & 2.3758 & 2.0048 &  2.3429 & $\bm{1.8509}$\\
\hline
\hline
\multicolumn{5}{c}{}\\
\hline
\hline

\multicolumn{5}{c}{\textbf{Barcelona Night (NN distortion)} from MPI-LFA}\\
\hline
\hline

\multicolumn{5}{c} { \adjustbox{valign=c,margin=0 1pt 0 1pt}{\includegraphics[width=0.98\columnwidth, height=0.28\columnwidth]{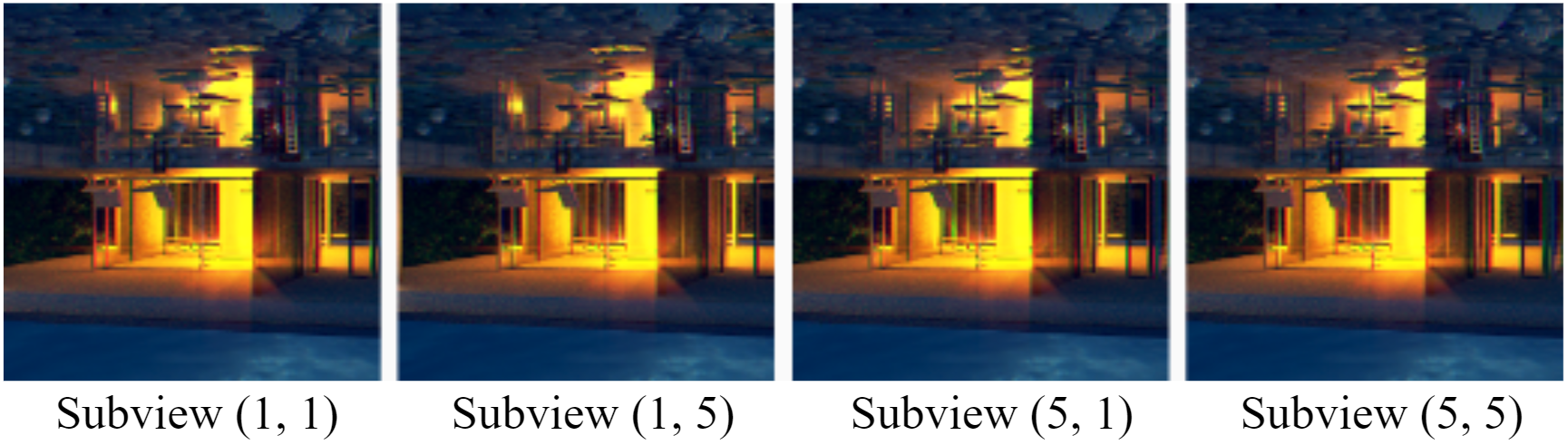}} }\\
\hline
\multicolumn{5}{c}{Ground Truth Quality Score: \underline{-0.2349}  (in [-9, 0])}\\
\hline
Metric & NR-LFQA & Tensor-NLFQ & ALAS-DADS & \textbf{LFACon}\\   
\hline
Quality Score  & -0.5605 & -1.3282 & -0.6537 & $\bm{-0.2040}$\\

\hline
\hline
\multicolumn{5}{c}{}\\
\hline
\hline

\multicolumn{5}{c}{\textbf{MSU0481 (SSDC distortion)} from SMART}\\
\hline
\hline

\multicolumn{5}{c} {\adjustbox{valign=c,margin=0 1pt 0 1pt}{\includegraphics[width=0.98\columnwidth, height=0.28\columnwidth]{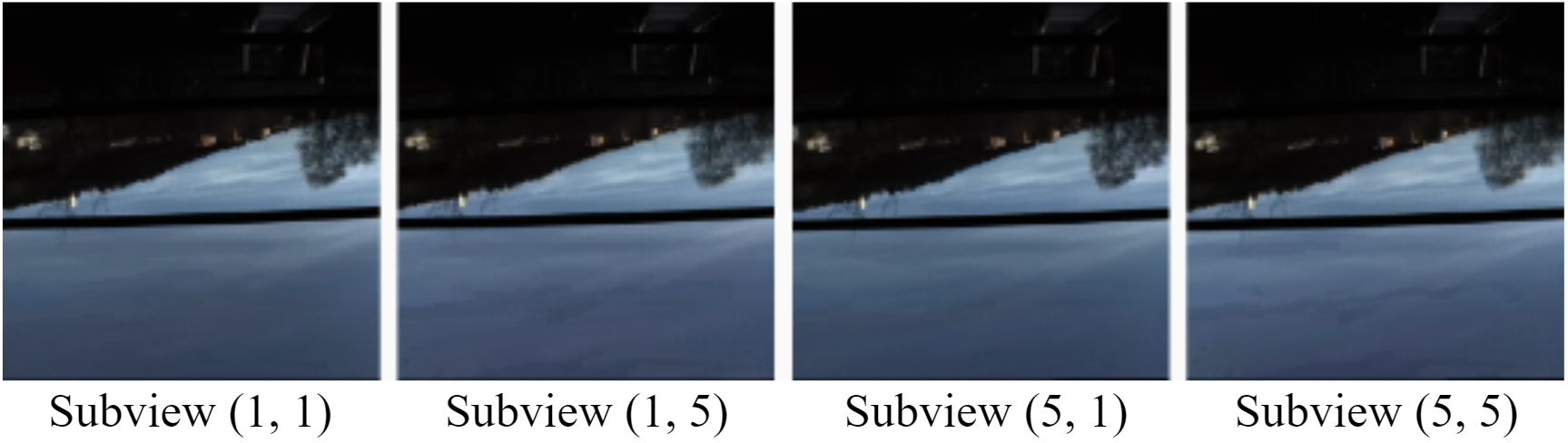}} }\\
\hline
\multicolumn{5}{c}{Ground Truth Quality Score: \underline{-2.4350}  (in [-13, 0])}\\
\hline
Metric & NR-LFQA & Tensor-NLFQ & ALAS-DADS & \textbf{LFACon}\\
\hline
Quality Score & -4.0230 & -3.6972 & -2.9612 & $\bm{-2.4344}$\\
\hline
\hline
\end{tabular}

\caption{Three LFIs are sampled from the three datasets and their subviews at angular positions (1, 1), (1, 5), (5, 1), and (5, 5) are shown. The predicted quality scores of LFACon (in bold) are contrasted with the ground truth (underlined) and the quality scores of the state-of-the-art LFIQA metrics. \textbf{The closer the predicted scores are to the ground truth, the more accurate the metrics are.}}
\label{fig:example_lfi}
\end{figure}

\subsubsection{Dataset Augmentation and Evaluation Metrics}
We augment the datasets by flipping the LFIs vertically and/or rotating them by 90, 180, and 270 degrees. As a result, the datasets are increased by a factor of eight, enlarging the Win5-LID dataset to 1760 LFIs, the SMART dataset to 2048 LFIs, and the MPI-LFA dataset to 2688 LFIs. This dataset augmentation approach not only helps in model training but also allows the model to better learn the image quality patterns \cite{qu2021light}.
We use the root mean square error (RMSE) \cite{dekking2005modern}, Spearman rank order correlation coefficient (SRCC) \cite{zwillinger1999crc}, Pearson linear correlation coefficient (PLCC) \cite{dekking2005modern}, and outlier ratio (OR) via Tukey's fences \cite{tukey1977exploratory} to assess the accuracy of an image quality assessment metric. For example, the output of a metric is more accurate or more consistent with the ground truth (human visual perception) when the RMSE values decrease and the SRCC or PLCC values increase.

\subsection{Experimental Results}
\subsubsection{Ablation Study for Pre- and Postattention Selection}
\label{subsec:ablation_study}
As mentioned earlier, we conduct an ablation study to determine the suitable pre- and postattention kernels for our proposed anglewise attention kernels. Specifically, four combinations of pre- and postattention kernels are examined: 1) linear neural projections for both (Linear + Linear); 2) $1 \times 1$ 3D convolutional layers for both (Conv3D + Conv3D); 3) a $1 \times 1$ 3D convolutional layer and an LF-DSC (Conv3D + LF-DSC); and 4) LF-DSCs for both (LF-DSC + LF-DSC). Their RMSE results are shown in Table \ref{tab:kernels}. We observe that the combination of ``Conv3D + LF-DSC'' achieves the best results for two out of the three datasets with a small number of parameters. Therefore, we adopt this combination as the default pre- and postattention kernels for our anglewise attention kernels, which are then integrated into our LFACon metric.

\subsubsection{Evaluation of Image Quality Assessment Performance}
\label{subsec:against_others}
To validate the efficacy of the proposed LFACon metric, we compare it with 12 full-reference image quality assessment metrics and 4 no-reference image quality assessment metrics, including 3 state-of-the-art no-reference LFIQA metrics, namely, NR-LFQA \cite{shi2019no}, Tensor-NLFQ \cite{zhou2020tensor}, and ALAS-DADS \cite{qu2021light}.
Table \ref{tab:benchmarking} displays the performance of the tested metrics on the Win5-LID, MPI-LFA, and SMART datasets. All the tested no-reference LFIQA metrics are supervised learning models, which are trained with the same training/testing ratio as that used for LFACon. As observed from Table \ref{tab:benchmarking}, LFACon achieves the best performance among all the benchmarked metrics. On Win5-LID, compared to the second-best model, LFACon gains impressive 28\%, 0.024, and 0.044 improvements in RMSE, SRCC, and PLCC, respectively. On the SMART dataset, LFACon achieves a 21\% improvement in RMSE over the second-best result. On the MPI-LFA dataset, LFACon yields a more substantial improvement, where it earns 50\%, 0.126, and 0.112 performance increases in RMSE, SRCC, and PLCC, respectively.
Overall, LFACon significantly outperforms all the state-of-the-art no-reference LFIQA models, including the recently proposed ALAS-DADS model. It should be noted that similar to LFACon, ALAS-DADS utilizes LF-DSC. The superior performance of LFACon over ALAS-DADS clearly demonstrates that the LFIQA performance can be noticeably improved by incorporating our proposed attention kernels.

Fig. \ref{fig:attn_maps} visualizes the workflow and the attention maps computed by the proposed anglewise attention kernels in LFACon for three sample LFIs. By directly observing the attention maps, it is difficult to see any discernible patterns in the attention weights, indicating the challenge in attention modeling. This validates the necessity of our attention modeling with state-of-the-art attention mechanisms. Our experimental results confirmed that the proposed anglewise attention mechanism provides sufficient feature extraction, feature concentration, and learning abilities in modeling such attention.

Fig. \ref{fig:example_lfi} shows some subjective results to qualitatively demonstrate the effectiveness of the proposed LFACon metric. We display three representative LFIs from the three datasets. Each LFI is distorted by a certain type and level of distortion and is represented by its four subviews in one row. In the meantime, we provide the quality score predicted by LFACon for each LFI compared to the state-of-the-art LFIQA metrics. These results show that the scores predicted by LFACon are much closer to the ground truth. For instance, the sample LFI ``Greek'' contains noticeable artifacts due to the JPEG2000 distortion. LFACon predicts its score as 1.8509, which is very close to the ground truth score of 1.8696. However, the other metrics, including NR-LFQA, Tensor-NLFQ, and ALAS-DADS, overestimated the quality at 2.3758, 2.0048, and 2.3429, respectively.

\begin{table*}[htbp]
\caption{RMSE/SRCC comparisons of competitive image quality assessment metrics for different distortion types. The best results are in bold. The rankings of LFACon are shown in the second-to-last column. The last column lists the improvement obtained by LFACon compared to the second-best metric (if LFACon is ranked first) or the difference compared to the best metric (if LFACon is not the best).}
\scriptsize
\renewcommand{\arraystretch}{1.2}
\setlength{\tabcolsep}{2pt}
\begin{tabular}{P{0.09\textwidth}|P{0.09\textwidth}|P{0.065\textwidth}P{0.065\textwidth}P{0.08\textwidth}P{0.08\textwidth}P{0.08\textwidth}P{0.09\textwidth}P{0.09\textwidth}P{0.08\textwidth}|P{0.04\textwidth}|P{0.05\textwidth}}
\hline
\hline
Distortion & Evaluation & PSNR & SSIM & MS-SSIM & BRISQUE & NR-LFQA & Tensor-NLFQ & ALAS-DADS& \textbf{LFACon} & \textbf{Rank}& \textbf{Boost}\\
\hline
\multirow{2}{*}{DQ} & RMSE $\downarrow$& 1.7402&  $1.8509$&  $1.8526$&  $1.6057$&  $1.0085$&  $1.2906$&  $1.4027$& $\bm{0.7739}$& $1$ & $+30\%$ \\ \cline{2-2}
& SRCC $\uparrow$&  $0.2474$&  $0.2378$&  $0.2302$&  $0.4694$&  $0.6002$&  $0.5099$&  $0.5109$& $\bm{0.8518}$& $1$ & $+0.25$ \\ \cline{1-12}
 \multirow{2}{*}{GAUSS} & RMSE $\downarrow$&  $1.5908$&  $1.6245$&  $1.6305$&  $1.4613$&  $0.8616$&  $1.0431$&  $0.9532$& $\bm{0.6113}$& $1$ & $+40\%$ \\ \cline{2-2}
& SRCC $\uparrow$&  $0.3462$&  $0.3195$&  $0.3276$&  $0.3200$&  $0.5458$& $\bm{0.6994}$&  $0.5024$&  $0.6359$& $2$ & $-0.06$ \\ \cline{1-12}
 \multirow{2}{*}{HEVC} & RMSE $\downarrow$&  $2.4933$&  $2.4766$&  $2.5510$&  $2.0591$&  $1.4641$&  $1.3750$&  $1.1699$& $\bm{0.6250}$& $1$ & $+87\%$ \\ \cline{2-2}
& SRCC $\uparrow$&  $0.6381$&  $0.7578$&  $0.7843$&  $0.2767$&  $0.5031$&  $0.5997$&  $0.7276$& $\bm{0.8370}$& $1$ & $+0.05$ \\ \cline{1-12}
 \multirow{2}{*}{LINEAR} & RMSE $\downarrow$&  $2.1339$&  $2.3727$&  $2.3605$&  $1.3453$&  $1.2970$&  $1.1480$&  $1.3792$& $\bm{0.8358}$& $1$ & $+37\%$ \\ \cline{2-2}
& SRCC $\uparrow$&  $0.4856$&  $0.4586$&  $0.5014$&  $0.4329$&  $0.7518$&  $0.8460$&  $0.8615$& $\bm{0.8793}$& $1$ & $+0.02$ \\ \cline{1-12}
 \multirow{2}{*}{NN} & RMSE $\downarrow$&  $1.4988$&  $1.5541$&  $1.5674$&  $0.8855$&  $0.7986$&  $1.3433$&  $1.3310$& $\bm{0.6885}$& $1$ & $+15\%$ \\ \cline{2-2}
& SRCC $\uparrow$&  $0.5683$&  $0.5466$&  $0.5809$&  $0.7243$&  $0.7898$&  $0.6039$&  $0.7291$& $\bm{0.8512}$& $1$ & $+0.06$ \\ \cline{1-12}
 \multirow{2}{*}{OPT} & RMSE $\downarrow$&  $1.6742$&  $1.7398$&  $1.7486$&  $1.5653$&  $1.2926$&  $1.3476$&  $1.2747$& $\bm{0.8281}$& $1$ & $+53\%$ \\ \cline{2-2}
& SRCC $\uparrow$&  $0.2472$&  $0.1980$&  $0.2442$&  $0.4168$&  $0.5407$&  $0.4829$&  $0.6722$& $\bm{0.6922}$& $1$ & $+0.02$ \\ \hline
\multirow{2}{*}{EPICNN} & RMSE $\downarrow$&
$1.3001$& $1.1649$& $1.0664$& $1.0193$& $\bm{0.1565}$& $0.2934$& $0.4448$&
 $0.2325$& $2$ & $-32\%$ \\ \cline{2-2}
& SRCC $\uparrow$& $0.7683$& $0.7450$& $0.8470$& $0.4909$& $\bm{0.9397}$&
$0.8624$& $0.6996$&  $0.9258$& $2$ & $-0.01$ \\ \cline{1-12}
 \multirow{2}{*}{JPEG2000} & RMSE $\downarrow$& $0.7857$& $0.8714$&
$0.8491$& $0.5773$& $0.3059$& $\bm{0.2976}$& $0.4450$&  $0.3781$& $3$ &
$-21\%$ \\ \cline{2-2}
& SRCC $\uparrow$& $0.7116$& $0.6018$& $0.6947$& $0.8344$& $\bm{0.9618}$&
$0.9501$& $0.9231$&  $0.9455$& $3$ & $-0.02$ \\ \cline{1-12}
 \multirow{2}{*}{USCD} & RMSE $\downarrow$& $0.8289$& $1.0166$& $0.9608$&
$0.7798$& $0.9931$& $1.1885$& $0.4553$& $\bm{0.3839}$& $1$ & $+18\%$ \\ \cline{2-2}
& SRCC $\uparrow$& $0.6702$& $0.5793$& $0.5560$& $0.8667$& $0.8126$&
$0.3426$& $0.9066$& $\bm{0.9243}$& $1$ & $+0.02$ \\ \hline
 \multirow{2}{*}{SSDC} & RMSE $\downarrow$&  $1.7447$&  $1.8537$&  $1.7757$&  $1.2179$&  $1.1728$&  $1.8321$&  $1.0344$& $\bm{0.7116}$& $1$ & $+45\%$ \\ \cline{2-2}
& SRCC $\uparrow$&  $0.6333$&  $0.3896$&  $0.4733$&  $0.7136$&  $0.7825$&  $0.7066$&  $0.7311$& $\bm{0.8715}$& $1$ & $+0.09$ \\ \hline  \hline
\end{tabular}
\label{tab:dist_bench}
\end{table*}
\begin{figure}[htb]
\centering
\subfloat[]{
    \includegraphics[width=\columnwidth]{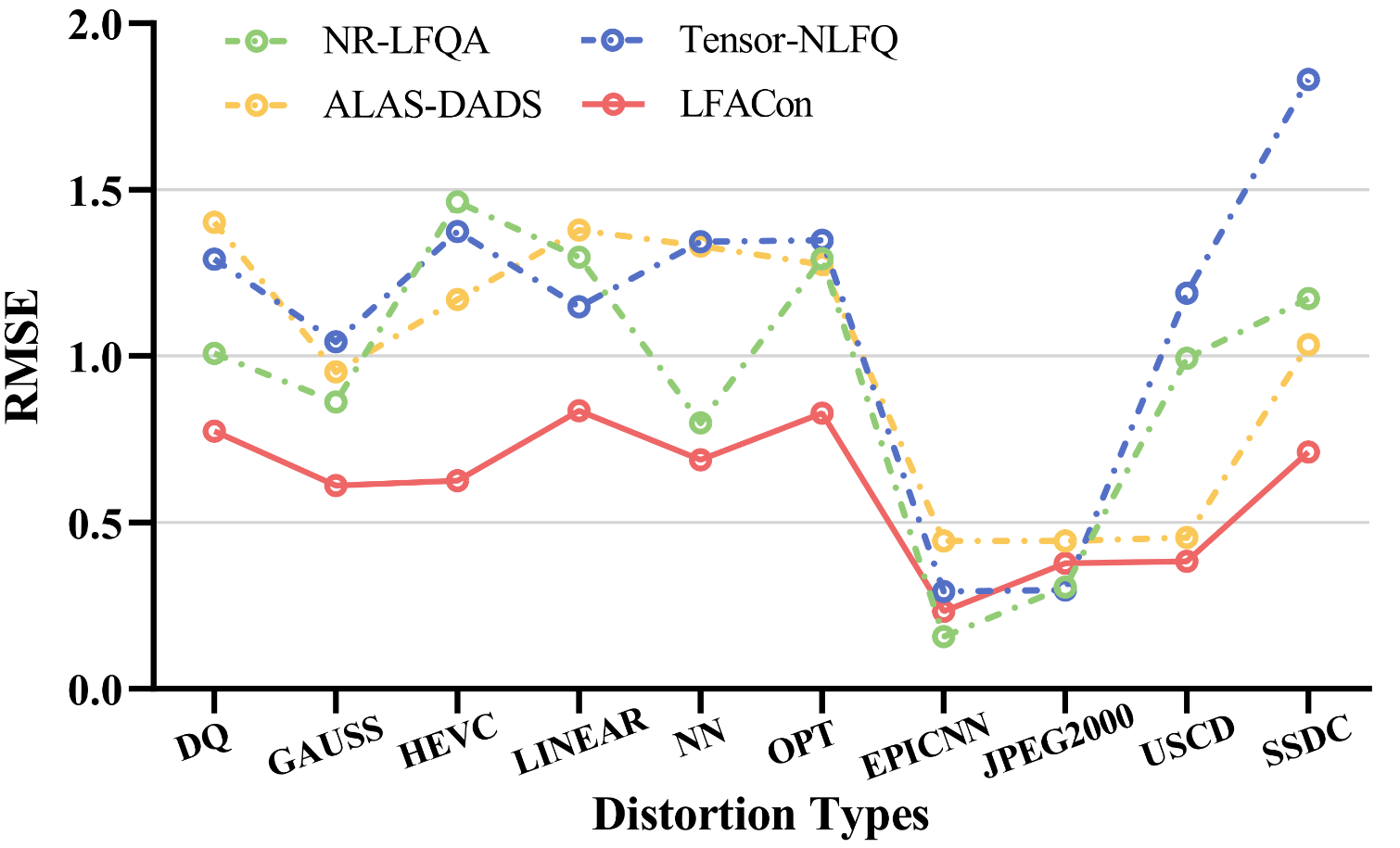}
}\\
\subfloat[]{
    \includegraphics[width=\columnwidth]{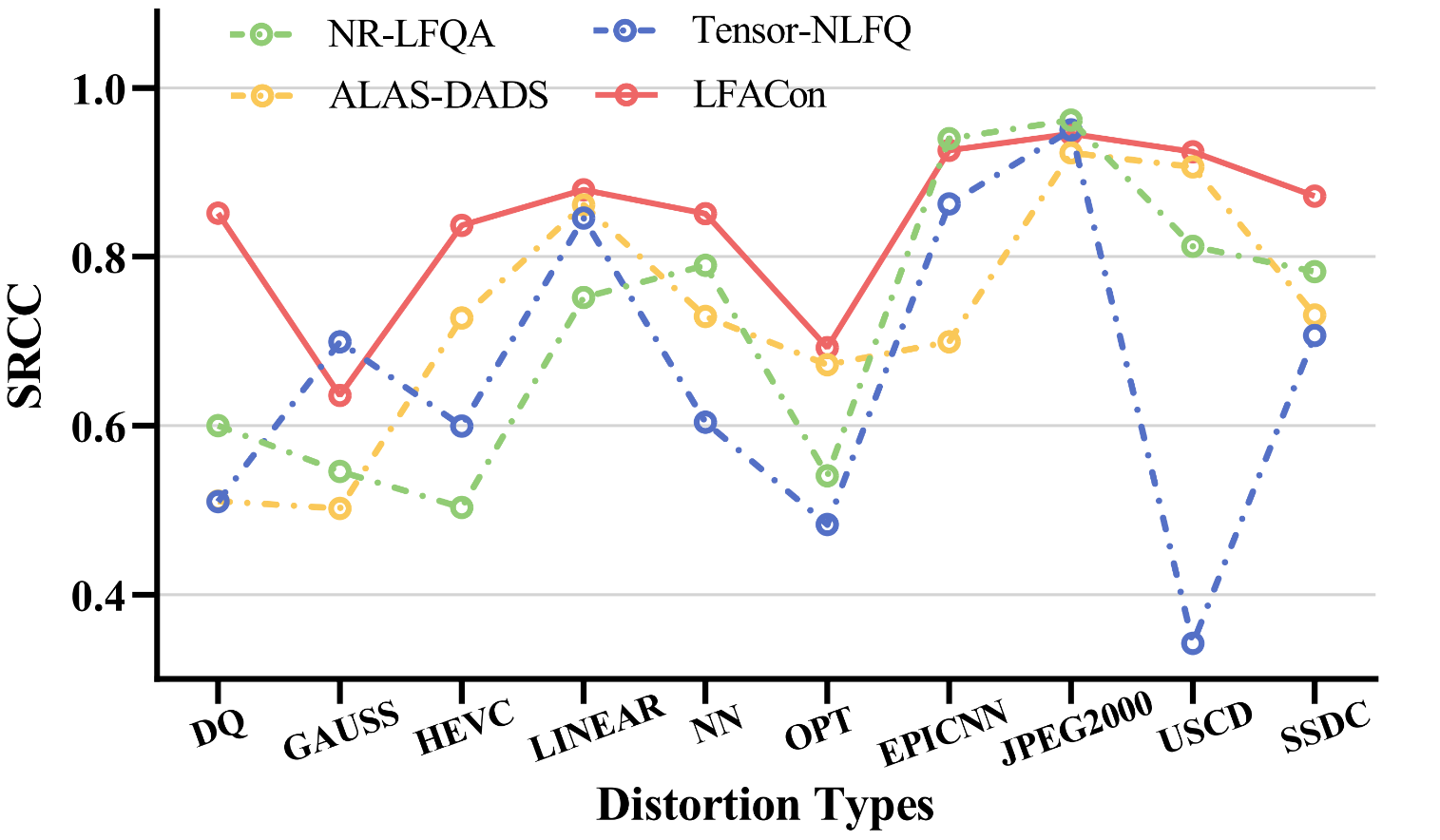}
}
\caption{Visualized comparisons of state-of-the-art LFIQA metrics
for different distortion types: (a) RMSE $\downarrow$ comparison. (b) SRCC $\uparrow$ comparison.}
\label{fig:RMSE_and_SRCC}
\end{figure}

\subsubsection{Effectiveness for Various Distortion Types}
\label{subsec:effective_distortion}

Table \ref{tab:dist_bench} demonstrates the performance of the evaluated image quality assessment metrics for various types of distortions.
We can observe from the table that LFACon's accuracy is ranked first for the majority of the distortion types. It achieves the lowest RMSE value for 8 out of the 10 distortion types and the greatest SRCC value for 7 distortion types. It is also noted that EPICNN is the distortion type for which NR-LFQA 
has superior performance. We believe this is because the NR-LFQA metric involves EPI-based processing, which may predominate its performance for EPICNN, which also employs EPI. In addition, LFACon shows significant advantages on the MPI-LFA dataset. For the distortion types of HEVC, OPT, and GAUSS, for instance, LFACon outperforms the second-best RMSE results by 87\%, 53\%, and 40\%, respectively. To provide a more intuitive view, Fig. \ref{fig:RMSE_and_SRCC} shows the RMSE/SRCC performance of LFACon in comparison to the state-of-the-art LFIQA metrics for different distortion types. It clearly demonstrates that LFACon achieves the best performance with the smallest RMSE and the largest SRCC for the majority of distortion types.

\begin{figure*}[htbp]
\centering
\includegraphics[width=\textwidth]{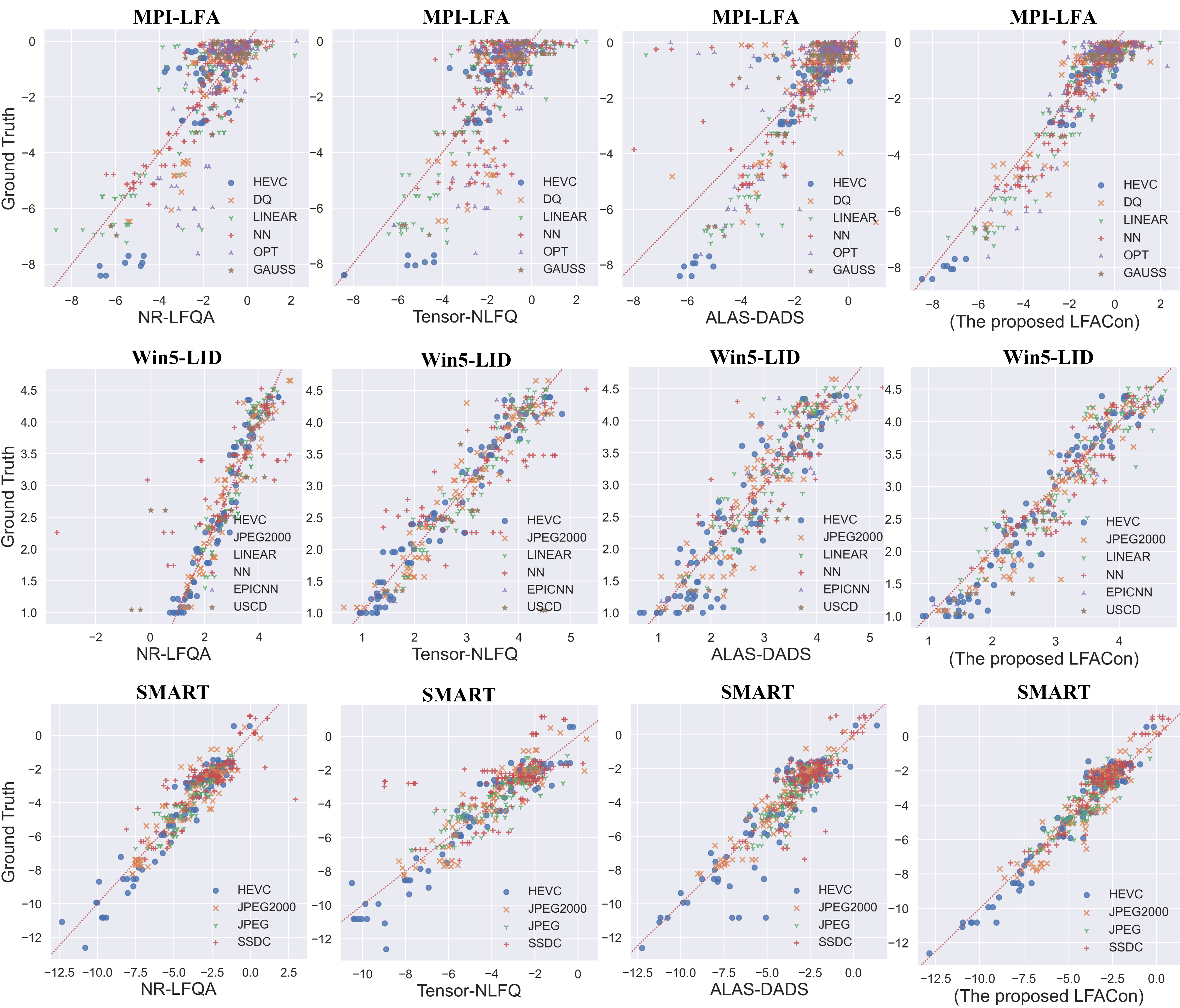}
\caption{Scatter plots of the ground truth image quality scores against the predicted quality scores by the tested no-reference LFIQA metrics on the MPI-LFA, Win5-LID, and SMART datasets. The distortion types are labeled with different symbols and colors. The red line in each subfigure shows the hypothetical ideal prediction line. \textbf{The closer the points are to the red line, the better the performance.}}
\label{fig:bench_plots}
\end{figure*}

Fig. \ref{fig:bench_plots} shows the scatter plots of the predicted image quality scores obtained by the tested no-reference LFIQA metrics and the ground truth quality scores for each distortion type on the MPI-LFA, Win5-LID, and SMART datasets. The red line in each subfigure represents the hypothetical ideal prediction line with a slope of 1, i.e., where the predicted quality scores are equal to the ground truth quality scores. By examining the subfigures for NR-LFQA, Tensor-NLFQ, and ALAS-DADS on the MPI-LFA dataset, for example, we find that these metrics yield biased predictions, i.e., underestimate the quality score when it is below
-2 (where 0 indicates the best quality) and overestimate the quality score otherwise.
LFACon, however, successfully minimizes such biased predictive patterns for all types of distortions. For example, Fig. \ref{fig:bench_plots} shows that LFACon's predicted quality scores are noticeably more centralized and closer to the hypothetical ideal prediction than those of the other metrics on all three datasets. This demonstrates that LFACon predicts not only with less bias but also with lower variance and significantly fewer outliers for all distortion types.

\begin{table}[htbp]
\scriptsize%
\centering
\renewcommand{\arraystretch}{1.2}
\caption{SRCC $\uparrow$ comparisons of LFACon predictions for different distortion levels.}
\setlength{\tabcolsep}{2pt}
\begin{tabular}{P{0.07\textwidth}|P{0.06\textwidth}P{0.06\textwidth}P{0.06\textwidth}P{0.06\textwidth}P{0.06\textwidth}P{0.06\textwidth}}
\hline
\hline
Distortion& Skip $1$ & Skip $4$ & Skip $7$ & Skip $10$ & Skip $17$ & Skip $24$ \\
\hline
DQ & $0.7015$ & $\bm{0.8334}$ & $0.5865$ & $0.5969$ & $0.7874$ & $0.4389$ \\
\hline
GAUSS & $0.1271$ & $0.6983$ & $\bm{0.7590}$ & $0.2206$ & $0.3591$ & $-0.8660$ \\
\hline
HEVC & $0.5988$ & $0.0684$ & $0.3015$ & $0.6357$ & $0.2052$ & $\bm{0.7807}$ \\
\hline
LINEAR & $0.3409$ & $0.6792$ & $0.2838$ & $0.3503$ & $0.6303$ & $\bm{0.7544}$ \\
\hline
NN & $0.4943$ & $0.7734$ & $\bm{0.8448}$ & $0.4869$ & $0.1017$ & $0.6955$ \\
\hline
OPT & $-0.0238$ & $0.5628$ & $0.3685$ & $0.4771$ & $0.0881$ & $\bm{0.8634}$ \\
\hline
\hline
\end{tabular}
\label{tab:self_dist}
\end{table}

\subsubsection{Variation at Different Distortion Levels}
\label{subsec:distortion_level}
\begin{figure*}[htbp]
\centering
\includegraphics[width=1\linewidth]{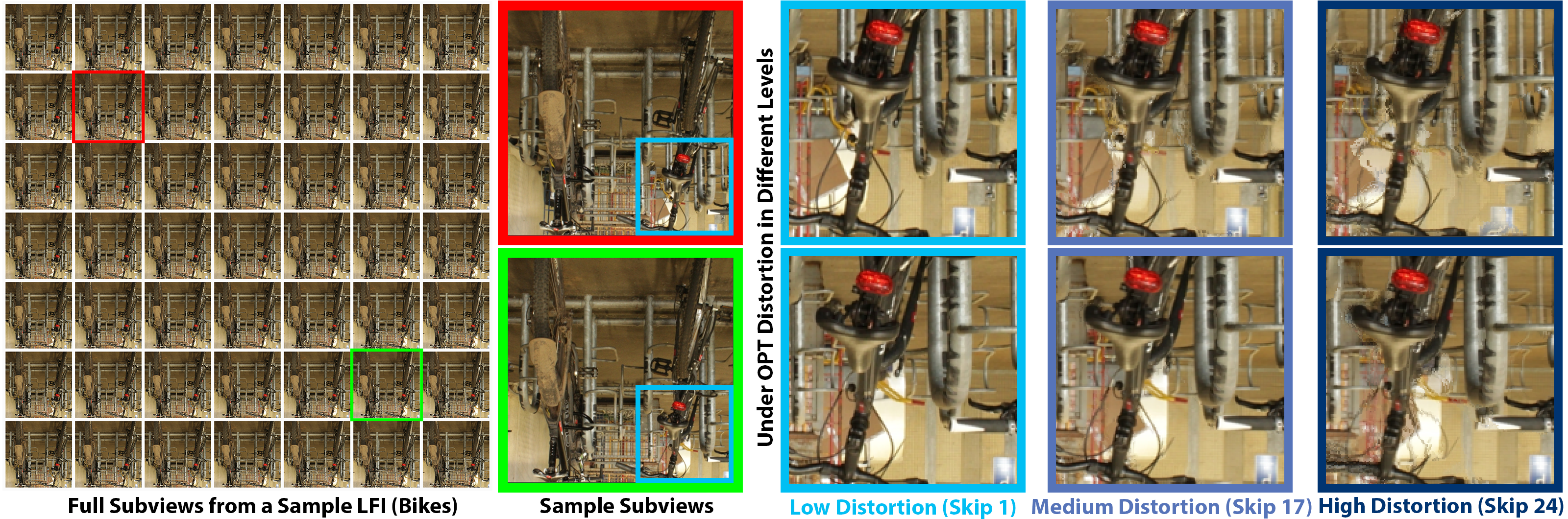}
\caption{A sample LFI (Bikes) from MPI-LFA showing the details of the subviews from two angular positions, followed by six zoomed-in images, that demonstrate the effects of low (skip 1), medium (skip 17), and high (skip 24) distortion levels.}
\label{fig:disto_level}
\end{figure*}

We are also interested in assessing the performance of LFACon at various distortion levels to learn more about its properties. To make the variations in successive distortion levels modest and comparable, the authors of the MPI-LFA dataset conducted a brief pilot study with 10 distortion levels and then manually selected the final levels \cite{kiran2017towards}. The final distortion level selection results for each kind, from moderate distortion to severe distortion, comprise skip 1, skip 4, skip 7, skip 10, skip 17, and skip 24. Fig. \ref{fig:disto_level} demonstrates how different distortion levels, e.g., low, medium, and high levels, affect the subviews from different viewpoints within the same scene. 

Table \ref{tab:self_dist} displays the SRCC results (not quality scores)  of LFACon at different distortion levels, where the bold values denote the highest SRCC value in each row (i.e., each distortion type). A higher SRCC value indicates that the predicted quality score is closer to the ground truth quality score. It is apparent that overall, LFACon works better at the low distortion levels of DQ, GAUSS, and NN, although it predicts more correctly at the high distortion levels (i.e., skip 24) of HEVC, LINEAR, and OPT. Even at the most severe distortion levels, LFACon is able to forecast with reasonable accuracy.

\begin{table}[htbp]
\scriptsize
\renewcommand{\arraystretch}{1.2}
\caption{Comparison of the computational time of the no-reference LFIQA metrics.}
\setlength{\tabcolsep}{2pt}
\begin{tabular}{P{0.09\textwidth}|P{0.06\textwidth}|P{0.055\textwidth}|P{0.055\textwidth}|P{0.055\textwidth}|P{0.055\textwidth}|P{0.055\textwidth}}
\hline
\hline
&\multicolumn{2}{c}{Win5-LID}&\multicolumn{2}{c}{SMART}&\multicolumn{2}{c}{MPI-LFA}\\
\hline
Metrics& Total (s)& Each (s) & Total (s)& Each (s)& Total (s)& Each (s) \\
\hline
NR-LFQA &  $3387$&  $9.622$& $4274$&  $10.450$&  $4797$& $8.933$ \\
Tensor-NLFQ &  $3085$&  $8.764$& $3441$&  $8.413$&  $4298$& $8.004$ \\
ALAS-DADS &  $313$&  $0.889$& $369$&  $0.902$&  $375$& $0.885$ \\
\textbf{LFACon} &  $\bm{218}$&  $\bm{0.619}$& $\bm{247}$&  $\bm{0.604}$&  $\bm{326}$& $\bm{0.607}$ \\
\hline
\textbf{Improvement} &  \multicolumn{2}{c|}{$\bm{+43.6\%}$}& \multicolumn{2}{c|}{$\bm{+49.4\%}$}&  \multicolumn{2}{c}{$\bm{+15.0\%}$} \\
\hline
\hline
\end{tabular}
\label{tab:pred_time}
\end{table}

\subsubsection{Evaluation of the Computational Efficiency}
\label{subsec:efficiency}
Table \ref{tab:pred_time} shows the computation times required by the tested no-reference LFIQA metrics to output the image quality scores on each dataset. We find that LFACon only requires approximately 0.6 seconds to predict the quality score of an LFI. This is on average approximately 36\% faster than the state-of-the-art ALAS-DADS metric and much faster than NR-LFQA and Tensor-NFLQ, both of which need more than 8 seconds to predict a quality score.

\subsubsection{Discussion}
\label{subsec:discussion}
Our experimental results show that the proposed LFACon metric outperforms both the traditional image quality assessment metrics and the state-of-the-art no-reference LFIQA metrics. We observe that for the majority of distortion types, LFACon achieves the highest accuracy. We believe the superior performance of LFACon substantially benefits from the incorporation of the proposed anglewise attention kernels. As discussed in Section \ref{subsec:aa}, these kernels extract multiangled features concurrently from the light field, capturing the key information that reflects the angular quality of LFIs under different distortion types. In addition, since the proposed attention kernels enable selective concentration on the key subviews in feature extraction, the extracted features are further concentrated on the essential quality cues. Moreover, as discussed in Section \ref{subsec:aa}, the proposed attention kernels can be readily integrated with existing lightweight preattention and postattention kernels. Our experiments testing LFACon with the integrated kernels show that it is significantly faster than the benchmark metrics.

We believe that our research not only advances no-reference LFIQA but also provides a theoretical foundation for a wider range of LFI research. The novel concepts and framework proposed in this research, e.g., the anglewise attention kernels and the corresponding LFACon learning model, can be naturally extended to other LFI research, such as superresolution \cite{yeung2018light, chen2022deep} and depth estimation \cite{wang2015occlusion, han2021novel}. For example, in LFI superresolution, the proposed attention kernels can be adapted as a lightweight and generative LFI feature extractor, which may take advantage of the computed anglewise attention to help increase the resolution of LFIs. Furthermore, depth estimation for LFIs involves processing multiple subviews similar to LFIQA. In this case, LFACon can be adopted as a general learning framework to rearrange the incorporated attention kernels to fit the depth estimation task. At last, it is also feasible to adapt the proposed anglewise attention to other emerging 3D scene representations. For instance, to support free viewpoints, neural radiance fields \cite{mildenhall2021nerf} can synthesize novel views of a complex 3D scene by using a sparse set of input views. The proposed anglewise attention kernels could be tailored to neural radiance fields, for example, to model the attention of the input views or reflect the multiview consistency of the synthesized novel views.


\section{Conclusion}
In this paper, we propose the novel concept of anglewise attention and propose three specific anglewise attention kernels. The proposed kernels are verified to be able to effectively extract quality features by modeling self-attention in an LFI. Based on these novel kernels, we propose our LFACon metric for no-reference LFIQA. Extensive experiments are conducted, and the results show that LFACon outperforms the state-of-the-art metrics and achieves the best performance for the majority of distortion types with reduced computational time.

In our future work, we will explore a broad range of applications of the proposed kernels in other light field processing tasks, such as superresolution and depth estimation. We may also attempt to adapt the proposed anglewise attention mechanism to other 3D scene representations, such as neural radiance fields.

\acknowledgments{
This work was supported in part by Beijing Natural Science Foundation under Grant 4222003 and National Natural Science Foundation of China under Grant 62177001.}

\bibliographystyle{abbrv-doi}

\bibliography{template}

\begin{thebibliography}{10}

\bibitem{ba2016layer}
J.~L. Ba, J.~R. Kiros, and G.~E. Hinton.
\newblock Layer normalization.
\newblock {\em arXiv preprint arXiv:1607.06450}, 2016.

\bibitem{chen2018accurate}
J.~Chen, J.~Hou, Y.~Ni, and L.-P. Chau.
\newblock Accurate light field depth estimation with superpixel regularization
  over partially occluded regions.
\newblock {\em IEEE Transactions on Image Processing}, 2018.

\bibitem{chen2022deep}
Y.~Chen, G.~Jiang, M.~Yu, H.~Xu, and Y.-S. Ho.
\newblock Deep light field spatial super-resolution using heterogeneous
  imaging.
\newblock {\em IEEE Transactions on Visualization and Computer Graphics}, 2022.

\bibitem{date2019full}
M.~Date, M.~Isogai, and H.~Kimata.
\newblock Full parallax table top 3d display using visually equivalent light
  field.
\newblock In {\em 2019 IEEE Conference on Virtual Reality and 3D User
  Interfaces (VR)}, pp. 1297--1298. IEEE, 2019.

\bibitem{dekking2005modern}
F.~M. Dekking, C.~Kraaikamp, H.~P. Lopuha{\"a}, and L.~E. Meester.
\newblock {\em A Modern Introduction to Probability and Statistics:
  Understanding why and how}.
\newblock Springer Science \& Business Media, 2005.

\bibitem{dosovitskiy2020image}
A.~Dosovitskiy, L.~Beyer, A.~Kolesnikov, D.~Weissenborn, X.~Zhai,
  T.~Unterthiner, M.~Dehghani, M.~Minderer, G.~Heigold, S.~Gelly, et~al.
\newblock An image is worth 16x16 words: Transformers for image recognition at
  scale.
\newblock {\em arXiv preprint arXiv:2010.11929}, 2020.

\bibitem{gonzalez2004fusion}
M.~Gonz{\'a}lez-Aud{\'\i}cana, J.~L. Saleta, R.~G. Catal{\'a}n, and
  R.~Garc{\'\i}a.
\newblock Fusion of multispectral and panchromatic images using improved ihs
  and pca mergers based on wavelet decomposition.
\newblock {\em IEEE Transactions on Geoscience and Remote sensing},
  42(6):1291--1299, 2004.

\bibitem{han2021novel}
K.~Han, W.~Xiang, E.~Wang, and T.~Huang.
\newblock A novel occlusion-aware vote cost for light field depth estimation.
\newblock {\em IEEE Transactions on Pattern Analysis and Machine Intelligence},
  2021.

\bibitem{huang2020light}
H.~Huang, H.~Zeng, Y.~Tian, J.~Chen, J.~Zhu, and K.-K. Ma.
\newblock Light field image quality assessment: An overview.
\newblock In {\em 2020 IEEE Conference on Multimedia Information Processing and
  Retrieval (MIPR)}, pp. 348--353. IEEE, 2020.

\bibitem{ihrke2016principles}
I.~Ihrke, J.~Restrepo, and L.~Mignard-Debise.
\newblock Principles of light field imaging: Briefly revisiting 25 years of
  research.
\newblock {\em IEEE Signal Processing Magazine}, 33(5):59--69, 2016.

\bibitem{itoh2016gaussian}
Y.~Itoh, T.~Amano, D.~Iwai, and G.~Klinker.
\newblock Gaussian light field: Estimation of viewpoint-dependent blur for
  optical see-through head-mounted displays.
\newblock {\em IEEE Transactions on Visualization and Computer Graphics},
  22(11):2368--2376, 2016.

\bibitem{jin2020light}
J.~Jin, J.~Hou, J.~Chen, S.~Kwong, and J.~Yu.
\newblock Light field super-resolution via attention-guided fusion of hybrid
  lenses.
\newblock In {\em Proceedings of the 28th ACM International Conference on
  Multimedia}, pp. 193--201, 2020.

\bibitem{kiran2017towards}
V.~Kiran~Adhikarla, M.~Vinkler, D.~Sumin, R.~K. Mantiuk, K.~Myszkowski, H.-P.
  Seidel, and P.~Didyk.
\newblock Towards a quality metric for dense light fields.
\newblock In {\em Proceedings of the IEEE Conference on Computer Vision and
  Pattern Recognition}, pp. 58--67, 2017.

\bibitem{koniaris2018compressed}
C.~Koniaris, M.~Kosek, D.~Sinclair, and K.~Mitchell.
\newblock Compressed animated light fields with real-time view-dependent
  reconstruction.
\newblock {\em IEEE Transactions on Visualization and Computer Graphics},
  25(4):1666--1680, 2018.

\bibitem{liu2021swin}
Z.~Liu, Y.~Lin, Y.~Cao, H.~Hu, Y.~Wei, Z.~Zhang, S.~Lin, and B.~Guo.
\newblock Swin transformer: Hierarchical vision transformer using shifted
  windows.
\newblock In {\em Proceedings of the IEEE/CVF International Conference on
  Computer Vision}, pp. 10012--10022, 2021.

\bibitem{lu2019improved}
Z.~Lu, H.~W. Yeung, Q.~Qu, Y.~Y. Chung, X.~Chen, and Z.~Chen.
\newblock Improved image classification with {4D} light-field and interleaved
  convolutional neural network.
\newblock {\em Multimedia Tools and Applications}, 78(20):29211--29227, 2019.

\bibitem{luong2015effective}
M.-T. Luong, H.~Pham, and C.~D. Manning.
\newblock Effective approaches to attention-based neural machine translation.
\newblock {\em arXiv preprint arXiv:1508.04025}, 2015.

\bibitem{lv20204d}
X.~Lv, X.~Wang, Q.~Wang, and J.~Yu.
\newblock 4d light field segmentation from light field super-pixel hypergraph
  representation.
\newblock {\em IEEE Transactions on Visualization and Computer Graphics},
  27(9):3597--3610, 2020.

\bibitem{ma2022arfnet}
S.~Ma, L.~Zhu, X.~Chen, X.~Yan, S.~Wang, P.~Yang, and B.~Xu.
\newblock Arfnet: Attention-oriented refinement and fusion network for light
  field salient object detection.
\newblock {\em IEEE Systems Journal}, 2022.

\bibitem{meng20203d}
X.~Meng, R.~Du, J.~F. JaJa, and A.~Varshney.
\newblock 3d-kernel foveated rendering for light fields.
\newblock {\em IEEE Transactions on Visualization and Computer Graphics},
  27(8):3350--3360, 2020.

\bibitem{mildenhall2021nerf}
B.~Mildenhall, P.~P. Srinivasan, M.~Tancik, J.~T. Barron, R.~Ramamoorthi, and
  R.~Ng.
\newblock Nerf: Representing scenes as neural radiance fields for view
  synthesis.
\newblock {\em Communications of the ACM}, 65(1):99--106, 2021.

\bibitem{mittal2012no}
A.~Mittal, A.~K. Moorthy, and A.~C. Bovik.
\newblock No-reference image quality assessment in the spatial domain.
\newblock {\em IEEE Transactions on Image Processing}, 21(12):4695--4708, 2012.

\bibitem{paudyal2017towards}
P.~Paudyal, F.~Battisti, M.~Sj{\"o}str{\"o}m, R.~Olsson, and M.~Carli.
\newblock Towards the perceptual quality evaluation of compressed light field
  images.
\newblock {\em IEEE Transactions on Broadcasting}, 63(3):507--522, 2017.

\bibitem{qu2021light}
Q.~Qu, X.~Chen, V.~Chung, and Z.~Chen.
\newblock Light field image quality assessment with auxiliary learning based on
  depthwise and anglewise separable convolutions.
\newblock {\em IEEE Transactions on Broadcasting}, 67(4):837--850, 2021.

\bibitem{ramachandran2017searching}
P.~Ramachandran, B.~Zoph, and Q.~V. Le.
\newblock Searching for activation functions.
\newblock {\em arXiv preprint arXiv:1710.05941}, 2017.

\bibitem{sheikh2006image}
H.~R. Sheikh and A.~C. Bovik.
\newblock Image information and visual quality.
\newblock {\em IEEE Transactions on image processing}, 15(2):430--444, 2006.

\bibitem{shi2019belif}
L.~Shi, S.~Zhao, and Z.~Chen.
\newblock {BELIF}: Blind quality evaluator of light field image with tensor
  structure variation index.
\newblock In {\em 2019 IEEE International Conference on Image Processing
  (ICIP)}, pp. 3781--3785. IEEE, 2019.

\bibitem{shi2018perceptual}
L.~Shi, S.~Zhao, W.~Zhou, and Z.~Chen.
\newblock Perceptual evaluation of light field image.
\newblock In {\em 2018 25th IEEE International Conference on Image Processing
  (ICIP)}, pp. 41--45. IEEE, 2018.

\bibitem{shi2019no}
L.~Shi, W.~Zhou, Z.~Chen, and J.~Zhang.
\newblock No-reference light field image quality assessment based on
  spatial-angular measurement.
\newblock {\em IEEE Transactions on Circuits and Systems for Video Technology},
  30(11):4114--4128, 2019.

\bibitem{tsai2020attention}
Y.-J. Tsai, Y.-L. Liu, M.~Ouhyoung, and Y.-Y. Chuang.
\newblock Attention-based view selection networks for light-field disparity
  estimation.
\newblock In {\em Proceedings of the AAAI Conference on Artificial
  Intelligence}, vol.~34, pp. 12095--12103, 2020.

\bibitem{tukey1977exploratory}
J.~W. Tukey et~al.
\newblock {\em Exploratory data analysis}, vol.~2.
\newblock Reading, MA, 1977.

\bibitem{vaswani2017attention}
A.~Vaswani, N.~Shazeer, N.~Parmar, J.~Uszkoreit, L.~Jones, A.~N. Gomez,
  {\L}.~Kaiser, and I.~Polosukhin.
\newblock Attention is all you need.
\newblock {\em Advances in neural information processing systems}, 30, 2017.

\bibitem{wald2000quality}
L.~Wald.
\newblock Quality of high resolution synthesised images: Is there a simple
  criterion?
\newblock In {\em Third conference" Fusion of Earth data: merging point
  measurements, raster maps and remotely sensed images"}, pp. 99--103.
  SEE/URISCA, 2000.

\bibitem{wang2015occlusion}
T.-C. Wang, A.~A. Efros, and R.~Ramamoorthi.
\newblock Occlusion-aware depth estimation using light-field cameras.
\newblock In {\em Proceedings of the IEEE international conference on computer
  vision}, pp. 3487--3495, 2015.

\bibitem{wang2002universal}
Z.~Wang and A.~C. Bovik.
\newblock A universal image quality index.
\newblock {\em IEEE Signal Processing Letters}, 9(3):81--84, 2002.

\bibitem{wang2004image}
Z.~Wang, A.~C. Bovik, H.~R. Sheikh, and E.~P. Simoncelli.
\newblock Image quality assessment: from error visibility to structural
  similarity.
\newblock {\em IEEE Transactions on Image Processing}, 13(4):600--612, 2004.

\bibitem{wang2003multiscale}
Z.~Wang, E.~P. Simoncelli, and A.~C. Bovik.
\newblock Multiscale structural similarity for image quality assessment.
\newblock In {\em The Thrity-Seventh Asilomar Conference on Signals, Systems \&
  Computers}, vol.~2, pp. 1398--1402. IEEE, 2003.

\bibitem{wanner2012globally}
S.~Wanner and B.~Goldluecke.
\newblock Globally consistent depth labeling of 4d light fields.
\newblock In {\em 2012 IEEE Conference on Computer Vision and Pattern
  Recognition}, pp. 41--48. IEEE, 2012.

\bibitem{wu2017light}
G.~Wu, B.~Masia, A.~Jarabo, Y.~Zhang, L.~Wang, Q.~Dai, T.~Chai, and Y.~Liu.
\newblock Light field image processing: An overview.
\newblock {\em IEEE Journal of Selected Topics in Signal Processing},
  11(7):926--954, 2017.

\bibitem{yang2007toward}
R.~Yang, X.~Huang, S.~Li, and C.~Jaynes.
\newblock Toward the light field display: Autostereoscopic rendering via a
  cluster of projectors.
\newblock {\em IEEE Transactions on Visualization and Computer Graphics},
  14(1):84--96, 2007.

\bibitem{yeung2018fast}
H.~W.~F. Yeung, J.~Hou, J.~Chen, Y.~Y. Chung, and X.~Chen.
\newblock Fast light field reconstruction with deep coarse-to-fine modeling of
  spatial-angular clues.
\newblock In {\em Proceedings of the European Conference on Computer Vision
  (ECCV)}, pp. 137--152, 2018.

\bibitem{yeung2018light}
H.~W.~F. Yeung, J.~Hou, X.~Chen, J.~Chen, Z.~Chen, and Y.~Y. Chung.
\newblock Light field spatial super-resolution using deep efficient
  spatial-angular separable convolution.
\newblock {\em IEEE Transactions on Image Processing}, 28(5):2319--2330, 2018.

\bibitem{yim2010quality}
C.~Yim and A.~C. Bovik.
\newblock Quality assessment of deblocked images.
\newblock {\em IEEE Transactions on Image Processing}, 20(1):88--98, 2010.

\bibitem{zhang2019self}
H.~Zhang, I.~Goodfellow, D.~Metaxas, and A.~Odena.
\newblock Self-attention generative adversarial networks.
\newblock In {\em International conference on machine learning}, pp.
  7354--7363. PMLR, 2019.

\bibitem{zhang2014vsi}
L.~Zhang, Y.~Shen, and H.~Li.
\newblock {VSI}: A visual saliency-induced index for perceptual image quality
  assessment.
\newblock {\em IEEE Transactions on Image processing}, 23(10):4270--4281, 2014.

\bibitem{zhou2020tensor}
W.~Zhou, L.~Shi, Z.~Chen, and J.~Zhang.
\newblock Tensor oriented no-reference light field image quality assessment.
\newblock {\em IEEE Transactions on Image Processing}, 29:4070--4084, 2020.

\bibitem{zwillinger1999crc}
D.~Zwillinger and S.~Kokoska.
\newblock {\em CRC standard probability and statistics tables and formulae}.
\newblock CRC Press, 1999.

\end{thebibliography}
\end{document}